\documentclass[aps,prb,twocolumn,floats]{revtex4-1}
\usepackage{graphicx}
\usepackage{amsmath}
\usepackage{amssymb}
\usepackage{bm}
\usepackage{amsmath}
\usepackage{mathtools}
\usepackage[normalem]{ulem}
\usepackage{color}

\begin{document}
\title{Topological transitions in a model with \\Particle-Hole symmetry, Pancharatnam-Berry Curvature and Dirac Points}
\author{P. V. Sriluckshmy}

\author{Archana Mishra}

\author{S. R. Hassan}

\author{R. Shankar}
\affiliation{The Institute of Mathematical Sciences, C.I.T. Campus, Chennai 600 113, India}

\date{\today}
\begin{abstract}
We study the topology and geometry of a fermionic model on the honeycomb lattice with spin-dependent hopping which breaks the time-reversal and charge-conjugation symmetries but preserves their composition. We show that in such a case the Zak phases are topological invariants that characterize the semi-metallic state at half-filling and determine the edge state structure. As the strength of the spin-dependent hopping varies, the model shows several Lifshitz transitions corresponding to creation and merging of multiple Dirac points. We discuss the possible realization of this model in cold atom systems and propose experimental signals of detecting the Dirac points and Pancharathnam-Berry curvature of the bands.
\end{abstract}
\maketitle
\section{Introduction}

There is much current interest in the topologically non-trivial phases of fermionic bands, since such phases have the potential of realizing quasi-particles with fractional quantum numbers and statistics. The topological phases of two dimensional insulators can be classified in terms of the sum of the Chern numbers of the occupied bands. The Chern numbers are the Pancharathnam-Berry (PB) curvature field integrated over the Brillioun zone. They are topological invariants as they do not change under smooth changes of the hamiltonian parameters that do not close the gap. The presence or absence of certain discrete symmetries leads to a more detailed classification of the possible phases \cite{schnyder2008,kitaev2009,lu2012}.

The PB curvature characterizes how the phases of the single-particle wave-functions twist over the Brillouin zone. The semi-classical Sundaram-Niu equations \cite{sundaram1999,haldane2004} provide a nice physical interpretation of PB curvature as magnetic field in momentum space. In the presence of an external force, it induces the so called anomalous velocity perpendicular to the external force. This leads to a Hall conductance even in the absence of an external magnetic field, a phenomenon dubbed the anomalous Hall effect. The value of the Hall conductance is equal to the PB curvature field integrated over the occupied states. When the highest occupied energy band is partially filled the system is called a topological Fermi liquid \cite{haldane2004}. When the Fermi level lies in the gap, the Hall conductance is the sum of the Chern numbers of the occupied bands and is hence quantized. Such systems are called Chern insulators.

The PB curvature becomes singular at the points where two bands touch due to the multivaluedness of the wave-function like at the Dirac points. Since the PB curvature is singular at the Dirac points, we refer to them as Dirac punctures (DP) in the Brillouin zone.  The Chern-number of each band becomes ill-defined but the sum of the Chern numbers of the two bands is still well defined. If two bands touch and detach at DPs as some hamiltonian parameter is varied, the Chern number can redistribute between the bands leading to a transition between two different topological phases. Several models with Dirac points \cite{hassan2013b,montambaux2009,hasegawa2012,zhu2007} and non-zero PB curvatures \cite{jo2012,tarruell2012,aidelsburger2011,jaksch2003,sorensen2005,gerbier2010,cooper2011,cooper2011b} have been studied.

In this paper we investigate the non-interacting limit of a model, the Kitaev-Hubbard model (KHUB) \cite{duan2003} presented by the hamiltonian, 
\begin{align}
\mathcal{H} & = -\sum_{\langle ij\rangle}C_{i\sigma}^{\dagger}
\frac{(tI +t'\sigma^{\alpha})_{\sigma \sigma'}}{2}C_{j\sigma'} + h.c.
+U\sum_in_{i\uparrow}n_{i\downarrow} \label{kh} 
\end{align}
describes a system of fermions on a honeycomb lattice. $i$ labels the sites of a honeycomb lattice, $\sigma$ the spin and $\alpha$ the nearest neighbour link. There is a spin-independent hopping term with strength $t$, a time-reversal breaking, spin-dependent hopping with strength $t'$ and an onsite repulsion term of strength $U$. 
\begin{figure}[ht]
 \begin{center}
 \includegraphics[width=7cm]{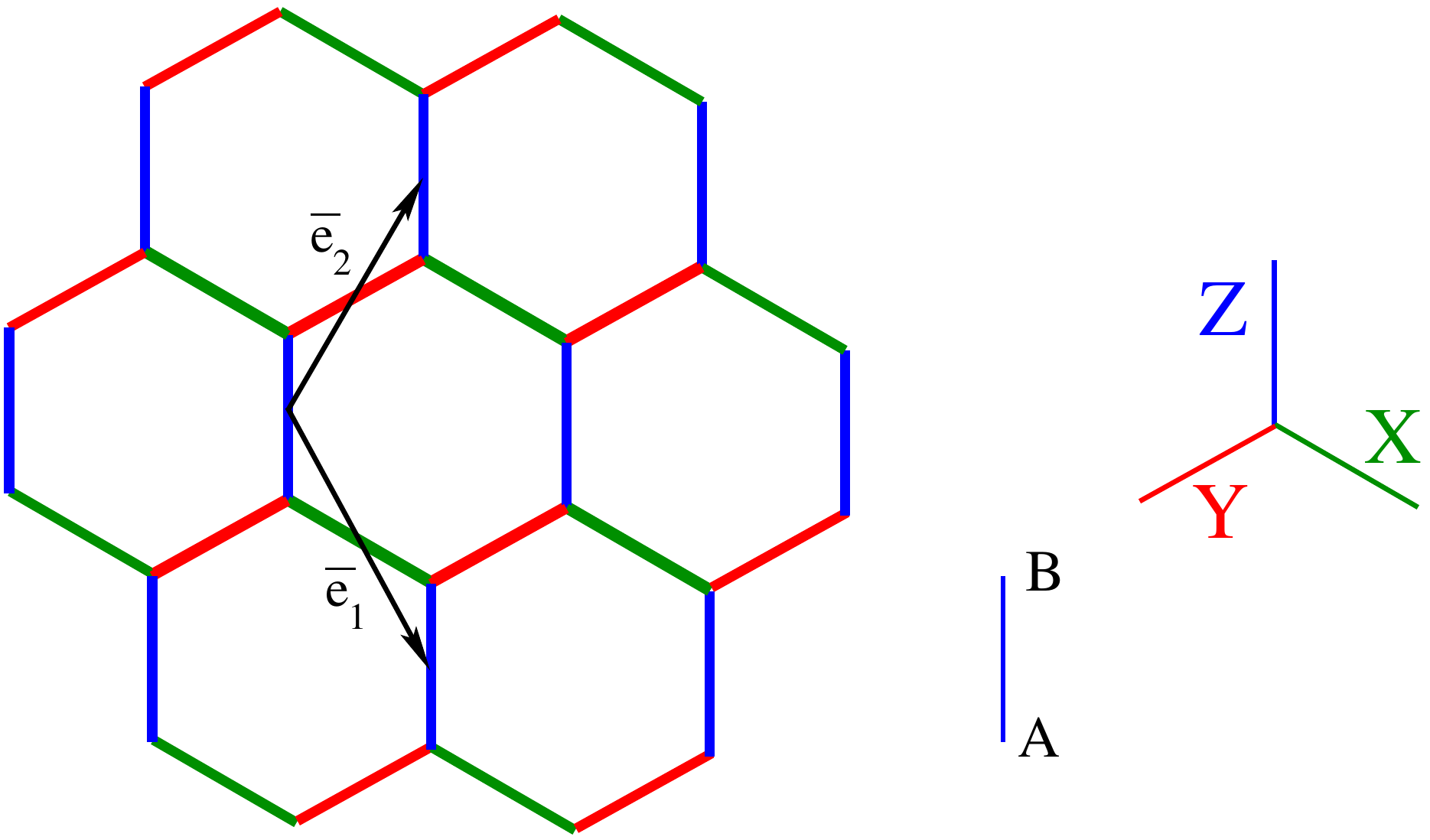}
 \caption{(color online) Honeycomb lattice has two sites per unit cell with basis vectors pointed along the $\vec{e}_1$ and $\vec{e}_2$ directions. The lower site is consistently taken to be the sublattice A and the upper sublattice B throughout this work.}
 \label{fig:honeycomb}
 \end{center}
 \end{figure}

 The KHUB model was originally proposed \cite{duan2003} as a way to realize Kitaev's honeycomb model \cite{kitaev2006} in the large $U/t$ limit, $t'/t=1$ and half filling.  The model has been studied earlier in the regime $1\ge t'/t\ge 0.5$ \cite{hassan2013,hassan2013b}. At half filling there is a chiral semi-metal phase at small $U$ and a Mott transition into an Algebraic Spin Liquid \cite{hassan2013} at large $U$.  At quarter filling there is a topological Fermi liquid phase and transition into Chern insulator phase \cite{hassan2013b}. The transition occurs at $t'/t=0.717$ at $U=0$. As $U$ is increased, the critical value of $t'/t$ decreases to $0.5$ at $U/t=10$.

We investigate the non-interacting limit, $U = 0$ in this work. There are topological transitions in this regime corresponding to creation and merging of DPs which occur at $t'/t = 0,1/\sqrt{3},\sqrt{3}$. It has been shown that these features persist even at non-zero values of $U$ \cite{jeanpaul2013}. We study, in detail, the topology of the phases and the transition between them. We also discuss the proposed experimental realization of the model in cold atom systems \cite{duan2003} and point out possible experimental signals of the topological features.

In section II, we review the effects of the discrete symmetries on the  PB curvature. We show that the PB phase vanishes for insulators with particle-hole symmetry at half filling. We then examine the topological features of bands with DPs and particle-hole symmetry in detail in section III. Here we explore the non-interacting hamiltonian of KHUB with periodic boundary condition (PBC) and show the topological transitions that characterize them. Further we study the model with open boundary conditions (OBC) and discuss the edge currents. In section IV we discuss the possible experimental realizations of the model and its topological features. Following the scheme proposed by Duan {\em et. al.}\cite{duan2003} we present a derivation of the spin-dependent periodic potential that leads to spin-dependent hopping terms in the hamiltonian. We show that the multiple DPs can be probed using Bloch-Zener oscillations and we present a semi-classical analysis to show how the PB curvature of the occupied bands can be detected. We summarise and discuss our results in section V.
\section{Discrete symmetries and the PB curvature}
As mentioned earlier, topological phases of insulators have been classified according to the presence or absence of certain discrete symmetries \cite{schnyder2008,kitaev2009,lu2012}, namely time reversal symmetry (TRS), charge conjugation symmetry (CCS) and their composition which we call particle-hole symmetry (PHS). In this section, these symmetries are briefly reviewed and the constraints that are imposed on the PB curvatures are examined for number-conserving, non-interacting, 2-dimensional fermionic systems. The hamiltonian for such systems can be written as  
\begin{equation}
\label{genham}
H = \int\frac{d^2k}{(2\pi)^2}~C^\dagger_a(k)h_{ab}(k)C_b(k)
\end{equation}
where $k$ goes over the Brillouin zone of a 2-dimensional Bravais lattice, $h=h^\dagger$ is the single-particle hamiltonian and $a,b=1,\dots,N_B$ label the sublattice and spin indices. The single-particle hamiltonian of KHUB in Eq.\eqref{kh}, has $N_B=4$ corresponding to two sublattice and two spin orbitals in every unit cell. Setting $t=1$, it can be written as
\begin{equation}
\label{khk}
h_{KHUB}(k,t')=\alpha^\dagger\otimes\Sigma(k,t')+\alpha\otimes\Sigma^\dagger(k,t')
\end{equation}
where $\alpha$ and $\Sigma$ are $2\times 2$ matrices in the sublattice and spin space respectively.
\begin{eqnarray}
\label{alphadef}
\alpha&=&\left(\begin{array}{cc}0&0\\1&0\end{array}\right)\\
\label{sigmadef}
\Sigma(k,t')&=&P_z+P_xe^{ik_1}+P_ye^{-ik_2}
\end{eqnarray}
where  $k_1 = \frac{1}{2}k_x -\frac{\sqrt{3}}{2}k_y$, $k_2 = \frac{1}{2}k_x +\frac{\sqrt{3}}{2}k_y$, $k_x = {\bf k}\cdot {\bf x},k_y = {\bf k}\cdot {\bf y}$ and $P_\alpha=\frac{1}{2}\left(I+t'\sigma^\alpha\right)$. We denote the spectrum as,
\begin{equation}
\label{genspect}
h(k)u^n(k)=\epsilon^n(k)u^n(k),~~n=1,\dots,N_B.
\end{equation}
In terms of these single particle eigen-functions, the PB vector potential, ${\cal A}^n_i(k)$ and curvature, ${\cal B}^n(k)$ are given by
\begin{equation}
{\cal A}^n_i(k)=i(u^n(k))^\dagger\frac{\partial u^n(k)}{\partial k_i},
~~{\cal B}^n(k)=\epsilon_{ij}\partial_i{\cal A}^n_j(k).
\end{equation}
From the PB curvature, the Chern number
\begin{align}
\nu_n & = \frac{1}{2\pi}\int \frac{d^2k}{4\pi^2}{\mathcal{B}}^n(k).
\end{align}
can be computed. Now we discuss effect of the discrete symmetries on the energy bands and the PB curvature, one by one.
\subsection{Time-reversal symmetry (TRS)}
The time-reversal transformation replaces particles (holes) with momentum $k$ by particles (holes) with momentum $-k$. It is an anti-unitary transformation which we denote by ${\cal T}$,
\begin{equation}
\label{trt}
{\cal T}C_a(k){\cal T}=\tau_{ab}C_b(-k),~~{\cal T}^\dagger{\cal T}=I,
~~{\cal T}i{\cal T}=-i
\end{equation}
All transition amplitudes are invariant under this transformation if there is a unitary matrix $\tau$ with $\tau^2=\pm1$ such that,
\begin{equation}
\label{trscond}
\tau^\dagger h^*(-k)\tau=h(k)
\end{equation}
Under the time-reversal transformation, ${\cal B}^n(k)=-{\cal B}^n(-k)$. Thus if it is a symmetry, then the Chern numbers, $\nu^n$ are all 0.

The KHUB satisfies the condition
\begin{equation}
\label{khtr}
h^*_{KHUB}(-k,t')=\sigma^y h(k,-t')\sigma^y .
\end{equation}
Thus for time-reversal symmetry to hold the condition in Eq.\eqref{trscond} needs to be satisfied for finite $t'$, implying that the matrix $\sigma^y\tau$ has to anti-commute with all the three Pauli matrices. Since such a matrix does not exist for any $t'$, the model in general is not TRS. But at two special points, $t'=0$ with $\tau=\sigma^y$ and $t'=\infty$ with $\tau=\beta\otimes\sigma^y$, where $\beta$ anti-commutes with $\alpha$ and $\alpha^\dagger$ the model preserves TRS.

\subsection{Charge conjugation symmetry (CCS)}

The charge-conjugation transformation replaces particles with momentum $k$ by holes with momentum $k$ and vice-versa. It is unitary transformation in the many-body Hilbert space that we denote by $\cal C$,
\begin{equation}
\label{cct}
{\cal C}C_a(k){\cal C}=\gamma_{ab}C^\dagger_b(-k),~~{\cal C}^\dagger{\cal C}=I,
~~{\cal C}i{\cal C}=i.
\end{equation}
All transition amplitudes are invariant under this transformation if there is a unitary matrix $\gamma$ with $\gamma^2=\pm1$ such that,
\begin{equation}
\label{ccscond}
\gamma^\dagger h^*(-k)\gamma=-h(k).
\end{equation}
If the system has CCS, then all the single particle energies come in pairs with $\epsilon^{\bar n}(k)=-\epsilon^n(k)$ and ${\cal B}^{\bar n}(k)=-{\cal B}^n(-k)$. $\bar n$ corresponds to the band index with negative of the energy of $n$. The positive and negative energy bands have opposite Chern numbers. From Eq.\eqref{khtr} it follows that the KHUB has CCS only at $t'=0$ (with $\gamma=\beta\otimes\sigma^y$) and at $t=0$ (with $\gamma=\sigma^y$).

\subsection{Particle-hole symmetry (PHS)}

The particle-hole transformation which we refer to as the composition ${\cal TC}$ replaces particles with momentum $k$ by holes with momentum $-k$ and vice versa. Note that the nomenclature is not uniform in the literature. For example Schnyder {\em et. al.} \cite{schnyder2008} refer to what we call CCS as the particle-hole symmetry and what we call PHS by ``chiral'' or ``sublattice'' symmetry. 

The particle-hole transformation defined above is a symmetry if
\begin{equation}
\label{phscond}
\gamma^\dagger\tau^\dagger h(k)\tau \gamma=-h(k)
\end{equation}
The KHUB has PHS with $\tau\gamma=\beta$ at all values of $t'$. This symmetry is very common in condensed matter systems. It occurs in all bipartite lattices where the fermion hopping is only from one sublattice to the other. PHS implies that all the single particle levels come in pairs with $\epsilon^{\bar n}(k)=-\epsilon^n(k)$ and ${\cal B}^{\bar n}(k)={\cal B}^n(-k)$. The sum of the PB curvature over the positive and negative energy bands are equal to zero individually as the sum of the PB curvature over all the bands is identically zero. Hence if there is PHS, then
\begin{equation}
\sum_{\epsilon^n(k)<0}{\cal B}^n(k)=0=
\sum_{\epsilon^n(k)>0}{\cal B}^n(k).
\end{equation}
Thus the total PB curvature vanishes for insulators with PHS at half-filling.
\section{Topology of bands with DP and PHS}
 In this section, we consider the case when the highest negative energy band and the lowest positive energy band touch at $N_D$ DPs, where $N_D$ is an even integer. We denote the DPs by ${\bf K}_n,~n=1,\dots N_D$. We will show that for systems with PHS at half filling the PB curvature is given by,
\begin{equation}
\label{dpvort}
{\cal B}(k)=\sum_{n=1}^{N_D}p_n\pi\delta^2({\mathbf k}-{\mathbf K_n})
\end{equation}
where $p_n$ is the PB flux passing through ${\mathbf K_n}$. Consequently, the Zak phases\cite{zak1989,saket2013,delplace2011}, $\Phi_Z$ defined as
\begin{equation}
\Phi_Z=\int_C {\cal A}_i(k)dk^i
\label{zakphase}
\end{equation}
are topological invariants. These quantities are independent of the contour $C$, provided it does not cross a DP. They are completely determined by the position and indices of the DPs. DPs lead to non-dispersive edge modes and we will show that the wave-vectors of these edge modes are determined by the Zak phases of the loops that wind around the Brillouin zone.

PHS implies that in the basis where $\beta$ is diagonal, the single-particle hamiltonian is of the form,
\begin{equation}
\beta=\left(\begin{array}{rr}I&0\\0&-I\end{array}\right),~~
h(k)=\left(\begin{array}{cc}0&\Sigma(k)\\\Sigma^\dagger(k)&0\end{array}\right).
\end{equation}
In general the two blocks defined above can have different dimensions, say $N$ and $M$, for example a bipartite lattice with different number of $A$ and $B$ lattice sites. However if $N\ne M$, there will be $\vert N-M\vert$ zero eigenvalues at every $k$, i.e. $\vert N-M\vert$ flat bands. While this may have interesting effects, we concentrate on the $N=M$ case so that we have an even number of bands, $N_B=2N$. It is then convenient to  replace the index $a=1,\dots,N_B$ by a pair $(r,\sigma),~r=A,B,~ \sigma=1,\dots,N$.

Using the fact that every matrix admits a singular value decomposition, we express $\Sigma$ as,
\begin{equation}
\label{svd}
\Sigma=U_A\epsilon U^\dagger_B
\end{equation}
where $U_{A(B)}$ are unitary matrices and $\epsilon$ is a diagonal matrix, $\epsilon_{nm}=\epsilon^n\delta_{nm},~\epsilon^n\ge 0, ~n,m=1,\dots,N$. The eigenvalues of the hamiltonian are then $\pm\epsilon^n$, the eigen-vectors being,
\begin{equation}
u^{\pm n}=\left(
\begin{array}{c}U^A\vert n\rangle\\\pm U^B\vert n\rangle\end{array}\right),
~~~~h \vert n\rangle=\epsilon^n\vert n\rangle.
\end{equation}
We can write $U_{A(B)}=e^{i\Omega_{A(B)}}\widetilde U_{A(B)}$, where $\widetilde U_{A(B)}$ are $SU(N)$ matrices with unit determinant. The PB vector potential and curvature summed over all the negative energy bands can be computed to be,
\begin{eqnarray}
{\cal A}_i(k)&=&\frac{1}{2}\partial_i\left(\Omega_A(k)-\Omega_B(k)\right) \label{vecpotential},\\
{\cal B}(k)&=&\frac{1}{2}\left(\partial_1\partial_2-\partial_2\partial_1\right)
\left(\Omega_A(k)-\Omega_B(k)\right)\label{berrycurv}.
\end{eqnarray}
Thus ${\cal B}(k)$ can be non-zero only at points where $\Omega(k)= \Omega_A(k)-\Omega_B(k)$ has a vortex type singularity. From Eq.\eqref{svd}, we see that $N\Omega$ is the phase of $\det\Sigma$. Since the matrix elements of $\Sigma$ are smooth functions of $k$, $\Omega$ can be multi-valued only at points where $\det\Sigma=0$. These are precisely the DPs. Thus we have proved Eq.\eqref{dpvort} showing that the PB curvature for systems with PHS at half filling is that of a set of vortices at the  DPs. We also see that $\det \Sigma(k)$ contains complete information of the topology of the system. The zeros of the determinant are the positions, ${\mathbf K}_n$, of the vortices. PB flux passing through ${\mathbf K_n}$ is $W_n\pi/N$, where $W_n$ is the winding number of the phase of ${\det}\Sigma(k)$ around it. We discuss the topological properties discussed above in the context of the KHUB in the following sections.

\subsection{Topology of KHUB with PBC}
We now apply the above developed formalism to the non-interacting hamiltonian KHUB which has PHS. It exhibits nontrivial properties \cite{hassan2013b} such as topological Lifshitz transitions and non-zero Chern numbers which we discuss in detail here, with $t$ fixed at unity.

\begin{figure}[ht]
 \begin{center}
 \includegraphics[width=6cm]{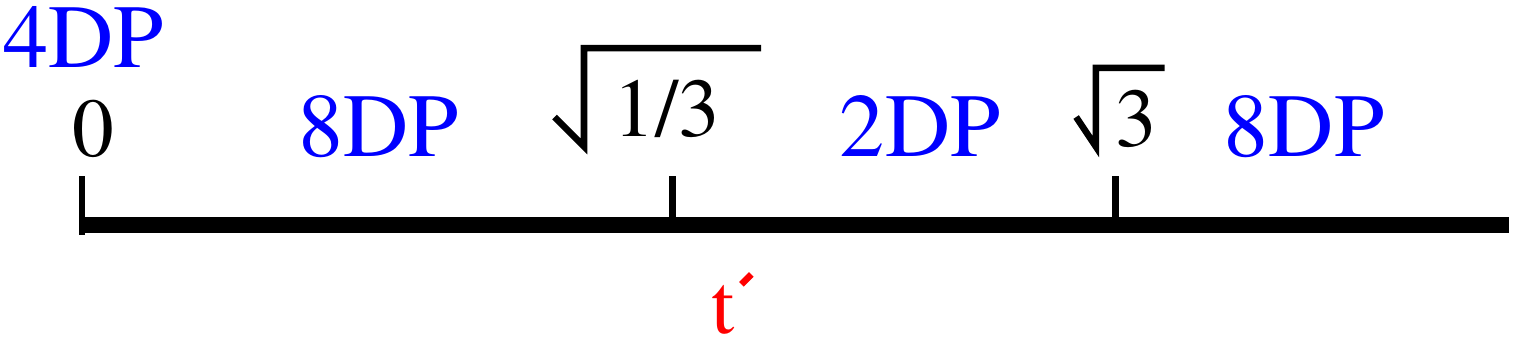}
 \caption{(color online) Number of DPs as a function of $t'$.}
 \label{fig:dp}
 \end{center}
 \end{figure}

The first and the second band of our four band hamiltonian overlap in the range $0\leq t'<0.717$ beyond which there is a non-zero gap between the bands at all $k$. Our model also features multiple DPs whose number changes as a function of the spin-dependent hopping parameter $t'$. The transition points are seen at $t'=0$, $1/\sqrt{3}$ and $\sqrt{3}$ (FIG.\eqref{fig:dp}). 
%
\begin{figure*}[ht]
 \begin{center}
 \includegraphics[width=4.5cm]{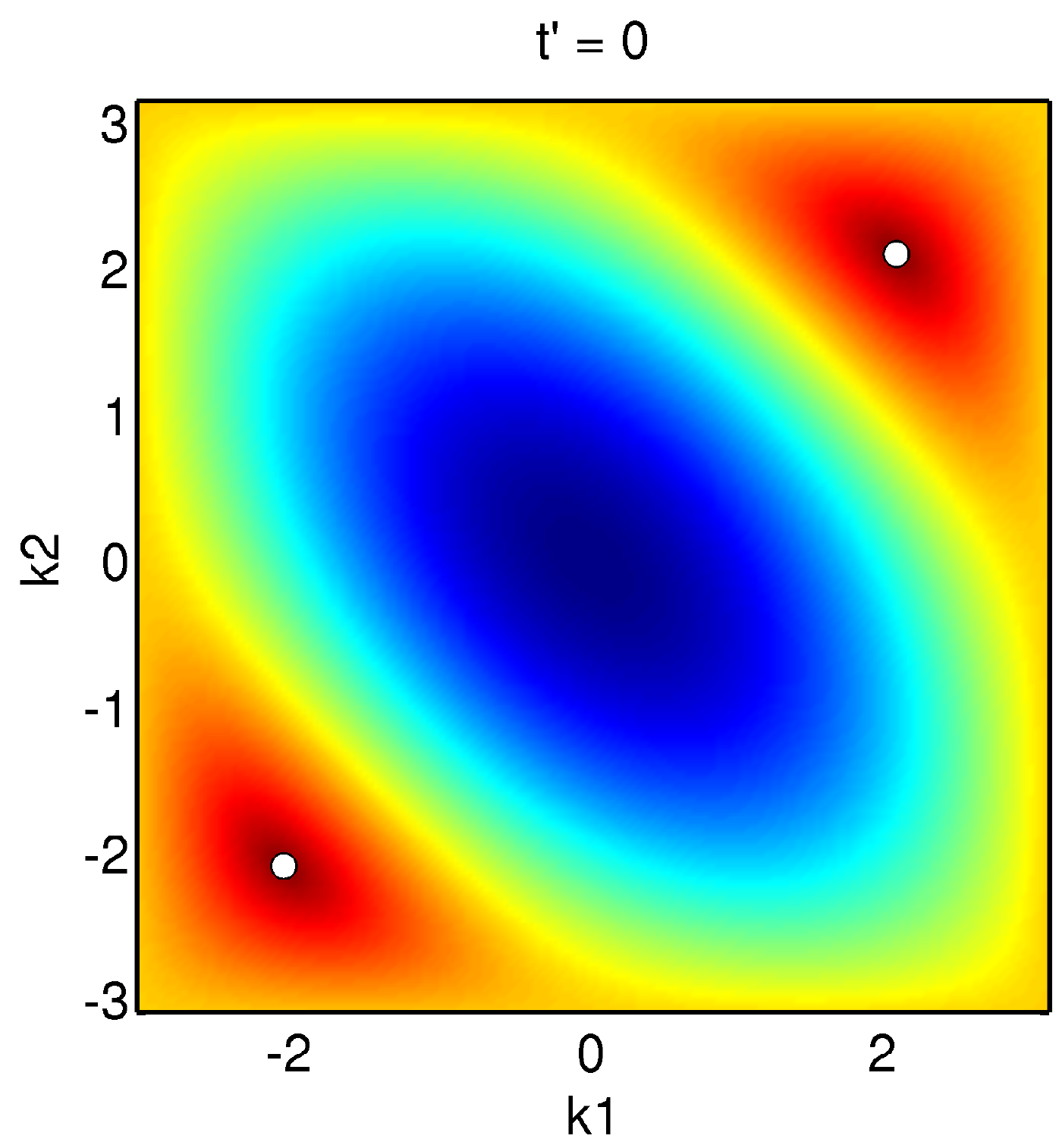}
\includegraphics[width=4.5cm]{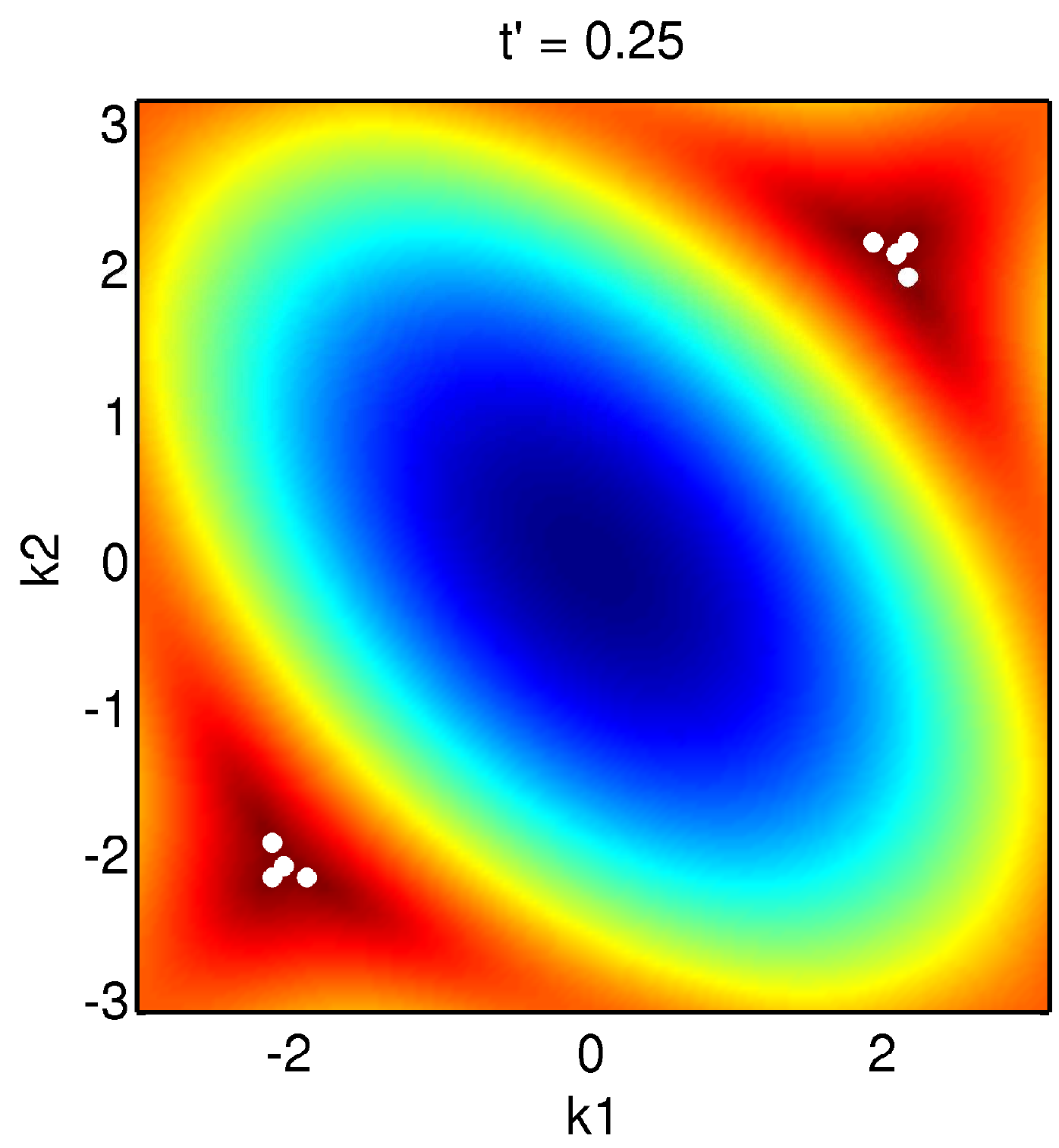}
\includegraphics[width=4.5cm]{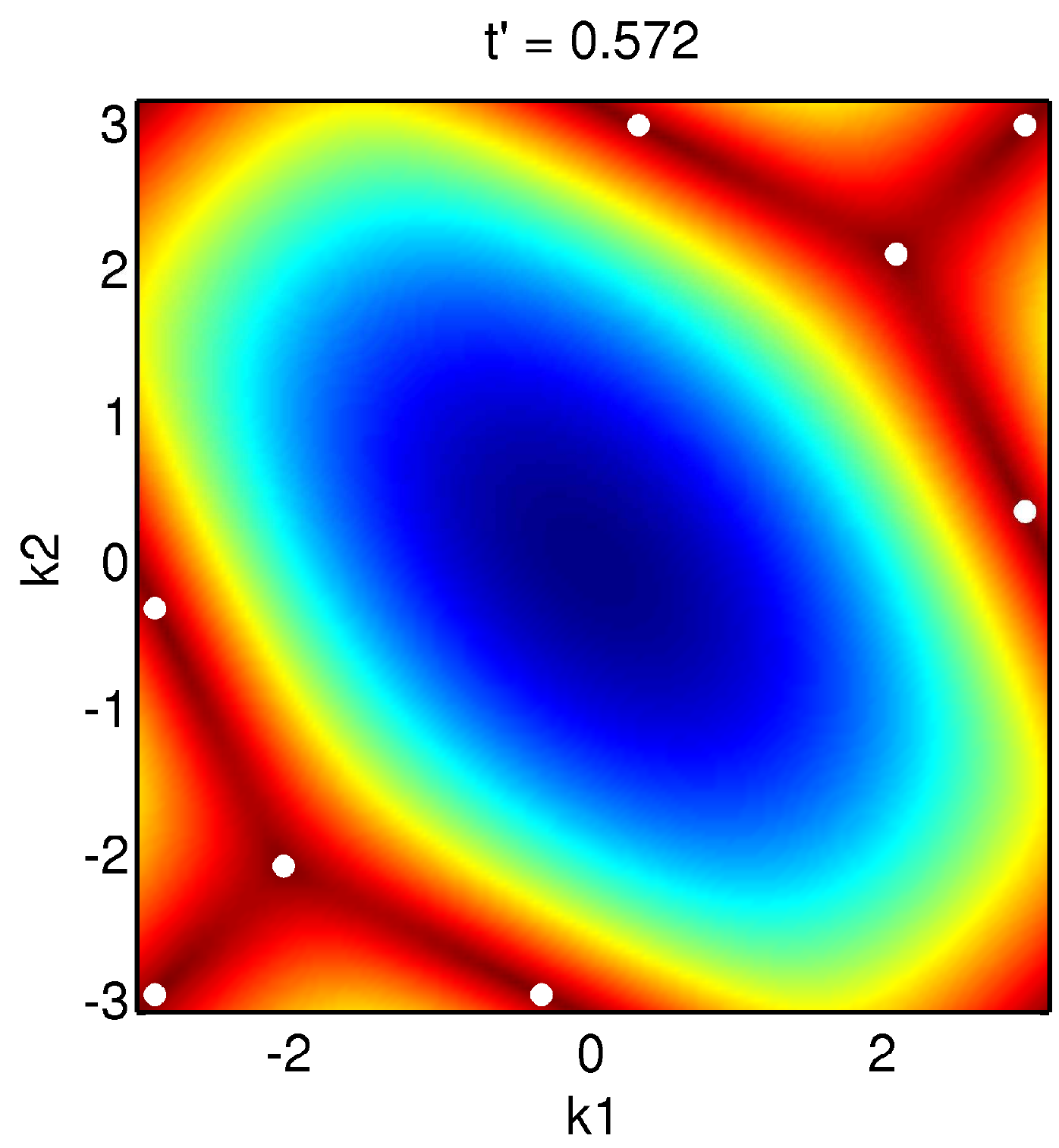}
\includegraphics[width=4.5cm]{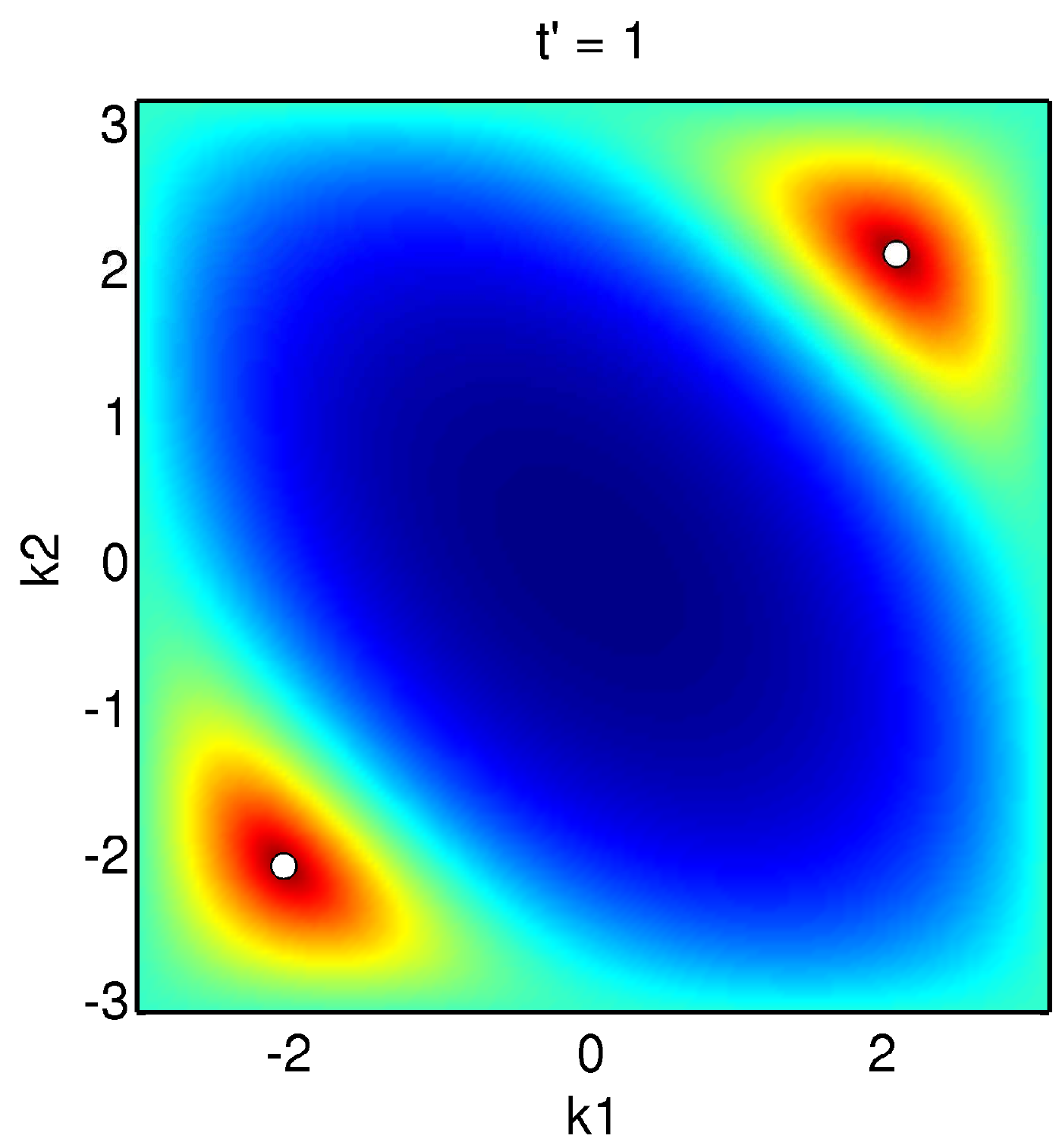}
\includegraphics[width=4.5cm]{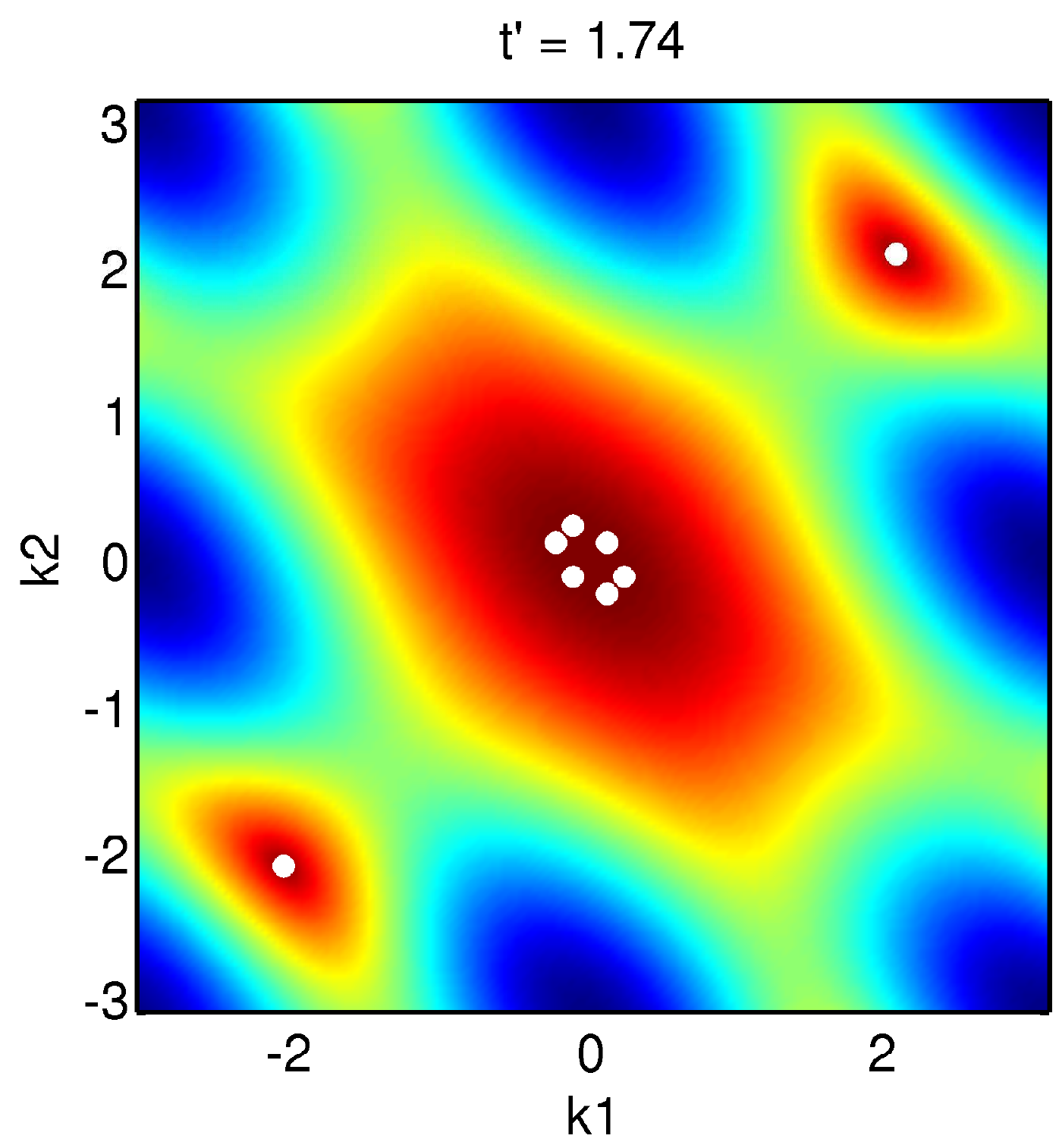}
\includegraphics[width=4.5cm]{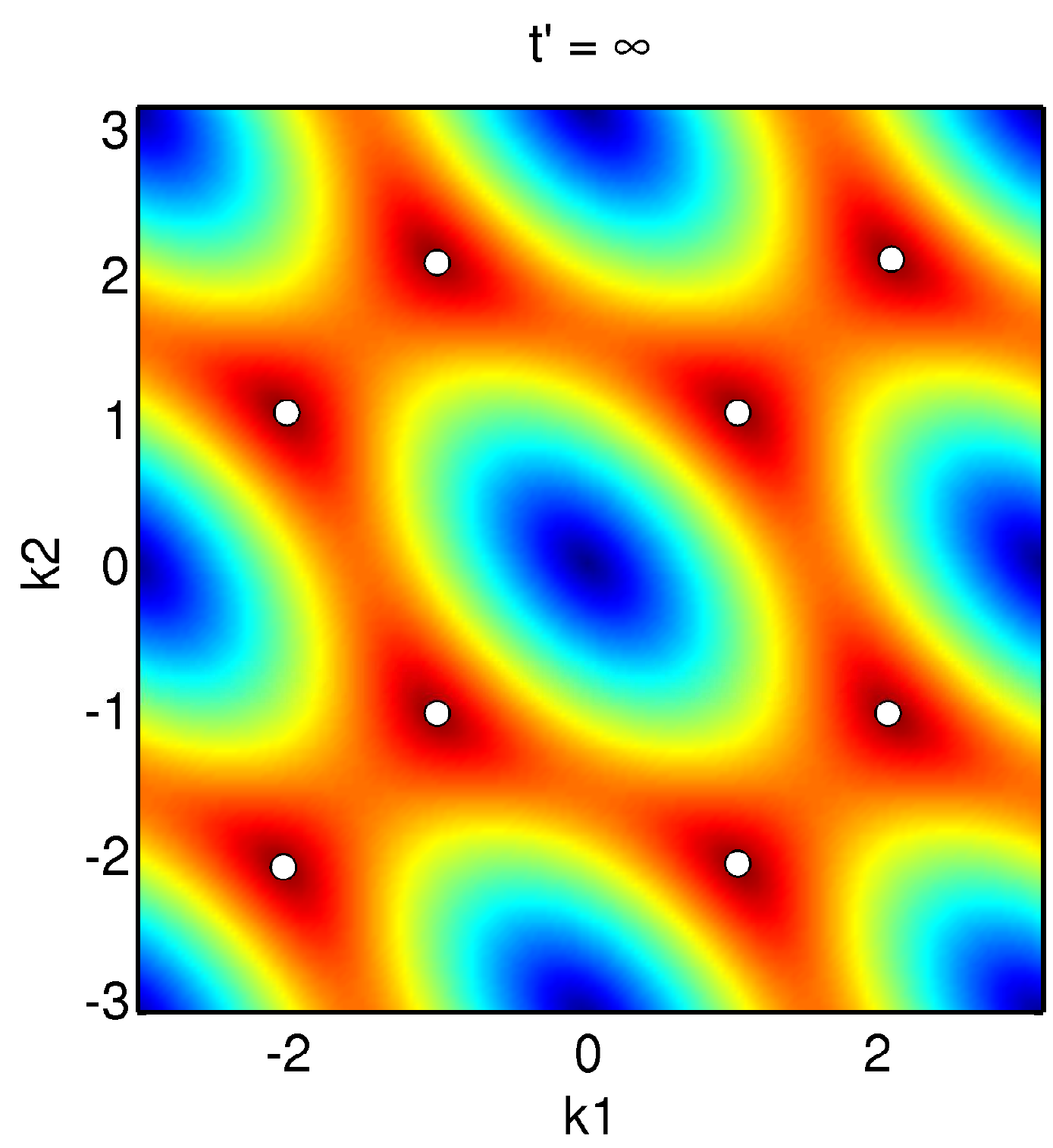}
 \caption{(color online) Pseudo color plots of the energy of the second band showing the DPs in the Brillouin zone for various $t'$.}
 \label{fig:energy_dp}
 \end{center}
 \end{figure*}
The location of the DPs can be determined from the energy spectra as the values at which the eigenvalues $\epsilon(k)$ vanishes. At the DPs the wave-functions of the two sublattices decouple and we get
\begin{align}
\Sigma(k,t') \psi_B & = 0 
\end{align}
We look for solutions in the $k_1 = k_2 = q$ direction, which imposes the condition on $q$ to be
\begin{align}
\pm t'\sqrt{1 + 2\cos(2q)} & = 1 + 2\cos q .\label{cond}
\end{align}
This condition is satisfied by $(q, q) = \pm {\bf K}_g = \pm (2\pi/3,2\pi/3)$ for all $t'$. At $t' = 0$, the graphene limit, doubly degenerate DPs are located at ${\bf K}_g $ and $-{\bf K}_g$ summing up to a total of $4$ DPs. With increasing $t'$, this condition is satisfied by another value of $q \in (0, \pi)$. Thus for $0<t'<1/\sqrt{3}$ there are a total of $8$ DPs located at $\pm {\bf K}$ given by
\begin{align}
{\bf{K}} = (2\pi /3,2\pi /3),(q,q),(q,2\pi -2q),(2\pi -2q,q) 
\end{align}
where the last two are related to $(q,q)$ through the underlying honeycomb lattice symmetry. At $t' = 1/\sqrt{3}$, six of these DPs merge in pairs at $(\pi,\pi)$, $(0,\pi)$ and $(\pi,0)$, leaving only those at $\pm{\bf{K}}_g$. For $t' \in (1/\sqrt{3},\sqrt{3})$, there are only 2DPs. At $t' = \sqrt{3}$, six DPs emerge from $(0,0)$ and move away from each other in the Brillouin Zone with increasing $t'$. FIG.\eqref{fig:energy_dp} shows the DPs for various $t'$. The merging and emerging of the DPs, previously discussed in other systems in \cite{degail2012,lim2012,fuchs2012,montambaux2009,wunsch2008}, is a topological Lifshitz transition \cite{degail2012,lim2012}.

In order to examine the Lifshitz transitions, we employ either the density of states or the thermodynamic consequences of the Fermi velocity, depending upon the transition point in question. The density of states does not change behaviour for the transition at $t'=0$. The Fermi velocity, which varies linearly with $t'$ for $t' > 0$ and is thus expected to vanish at $t'=0$, remains non-zero and finite at that value. This should reflect in many of the thermodynamic properties of the system, and thus it can be used as a probe of this Lifshitz transitions.

The energy dispersion relation of the system for $t' \in [0, 1/\sqrt{3})$ close to each of the DP is linear and is given by
\begin{align}
\epsilon & = \sqrt{a(t')q_1^2 + b(t')q_2^2}
\end{align}
where $q_1$ and $q_2$ are small deviations away from the DP and $a(t')$ and $b(t')$ are constants dependent on $t'$. The density of states $\rho(\omega)$ thus varies linearly with the energy $\omega$ for all $t'$ except for the values at which the DPs merge and emerge. At $t'=1/\sqrt{3}$, its behaviour changes sharply, with the dominant contribution varying as the square root of the energy. This is because, the dispersion relation takes the form 
\begin{align}
\epsilon & = \sqrt{a(t')q_1^2 + b(t')q_2^4}
\end{align}
at $(\pi,\pi)$, $(0,\pi)$ and $(\pi,0)$ merging points for $t'=1/\sqrt{3}$. On the other hand around the $(0,0)$ emerging point for $t'=\sqrt{3}$,  the dispersion relation takes the form
\begin{align}
\epsilon & = \sqrt{a(t')q_1^4 + b(t')q_2^4}.
\end{align}
thereby giving a constant and a linear contribution to the density of states. Very close to $\omega = 0$, however the constant term dominates. This sharp change in the density of states at $t' = 1/\sqrt{3}$ and $t' = \sqrt{3}$ makes it possible to probe the Lifshitz transitions.
 
\begin{figure}[ht]
 \begin{center}
 \includegraphics[width=4.2cm]{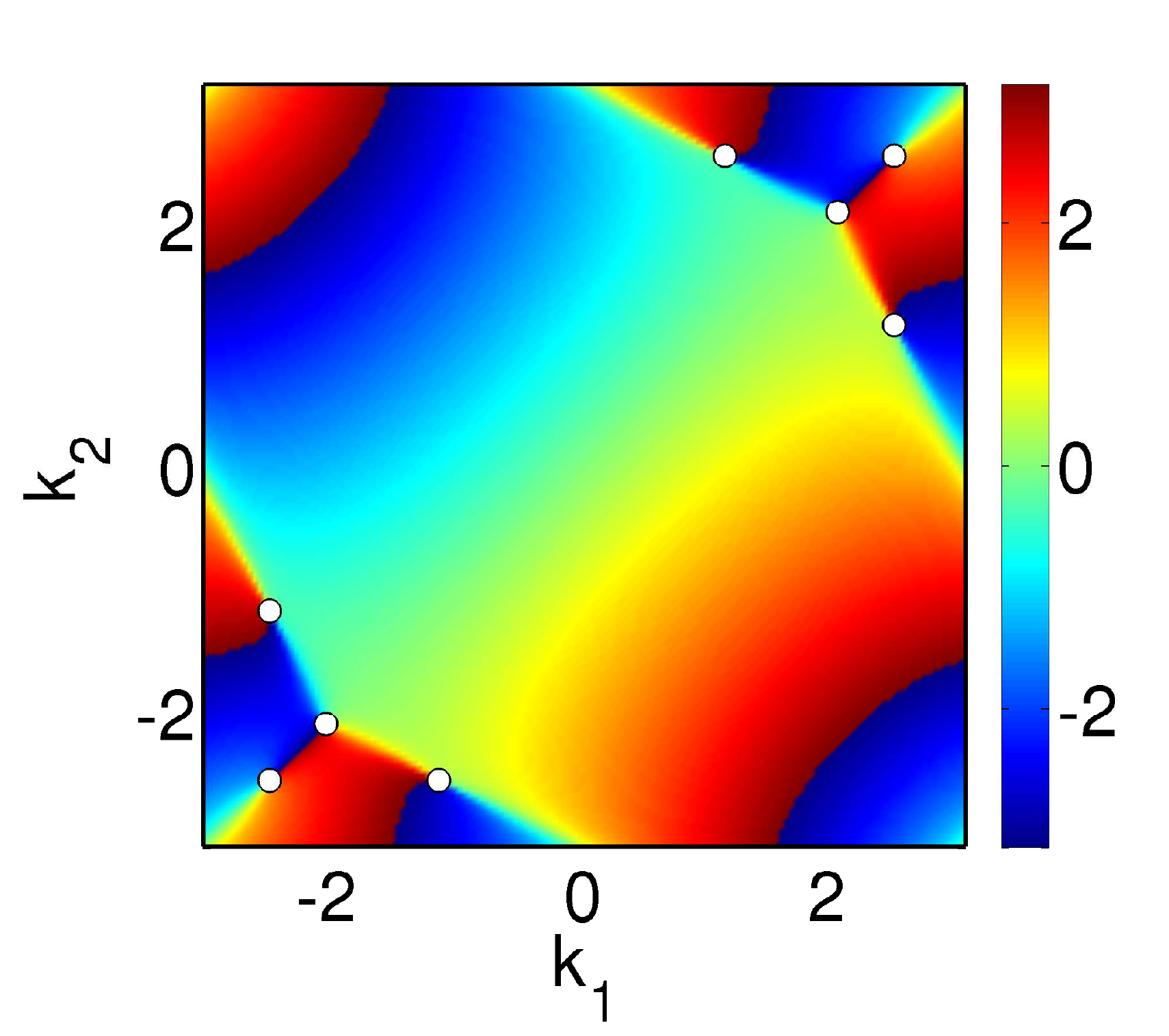}
 \includegraphics[width=4.2cm]{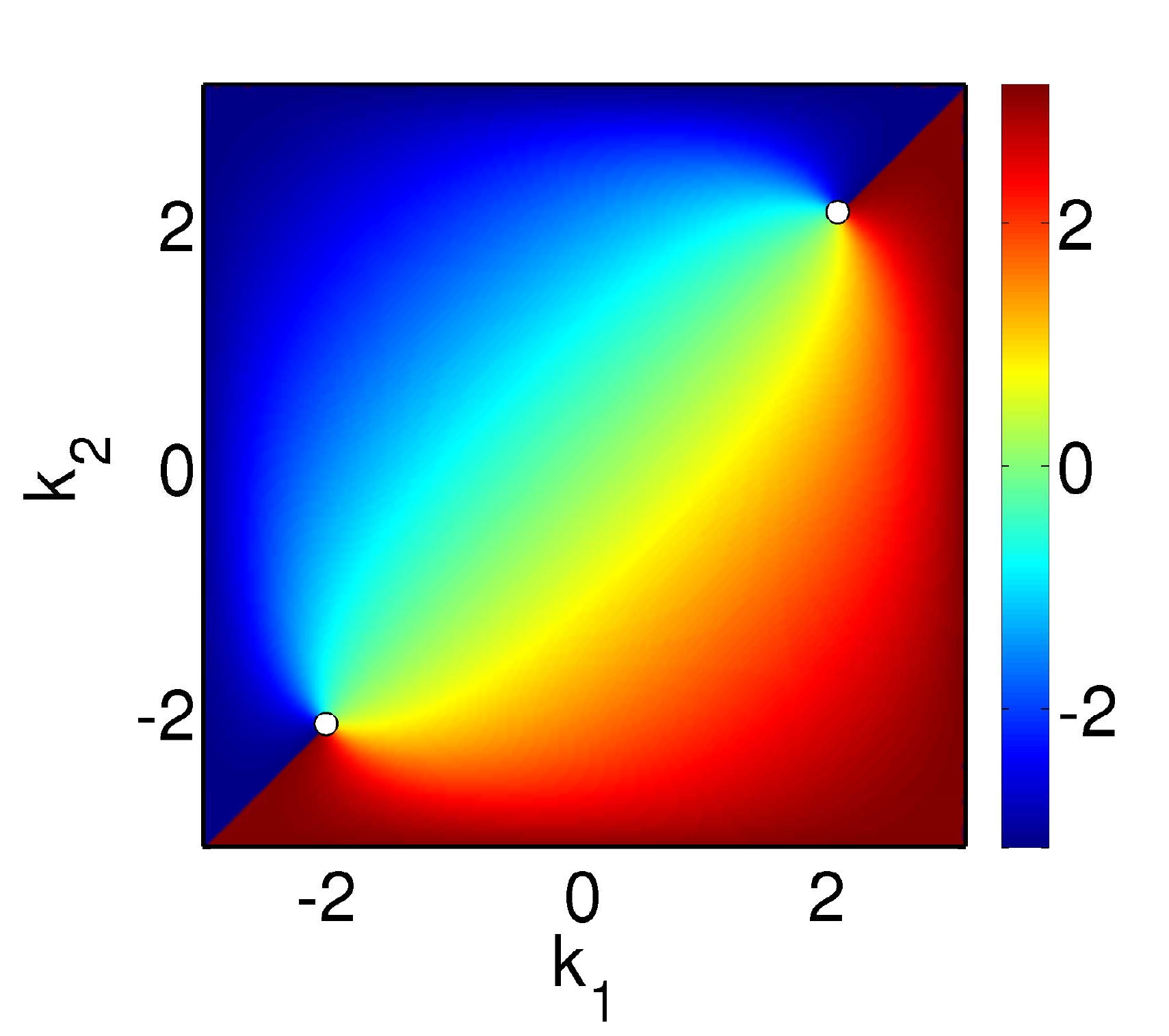}
 \caption{(color online) The phase of the determinant as a function of $(k_1,k_2)$ for $t' = 0.5$ and $t' = 1$. The phase changes discontinuously at the white dots which represents the location of the DPs.}
 \label{fig:determinant}
 \end{center}
 \end{figure}

Information about the DPs as well as the PB curvature of the system, as shown earlier, can be obtained from the phase of $\det\Sigma(k_1,k_2)$. In FIG.\eqref{fig:determinant}, we plot this for various values of $t'$. For $t' = 0.5$, there are eight distinct points around which the phase changes discontinuously by a value of $\pm 2\pi$ corresponding to the DPs. On the other hand, there are only two such points at $t' = 1$.
 
\begin{figure}[ht]
 \begin{center}
 \includegraphics[width=7cm]{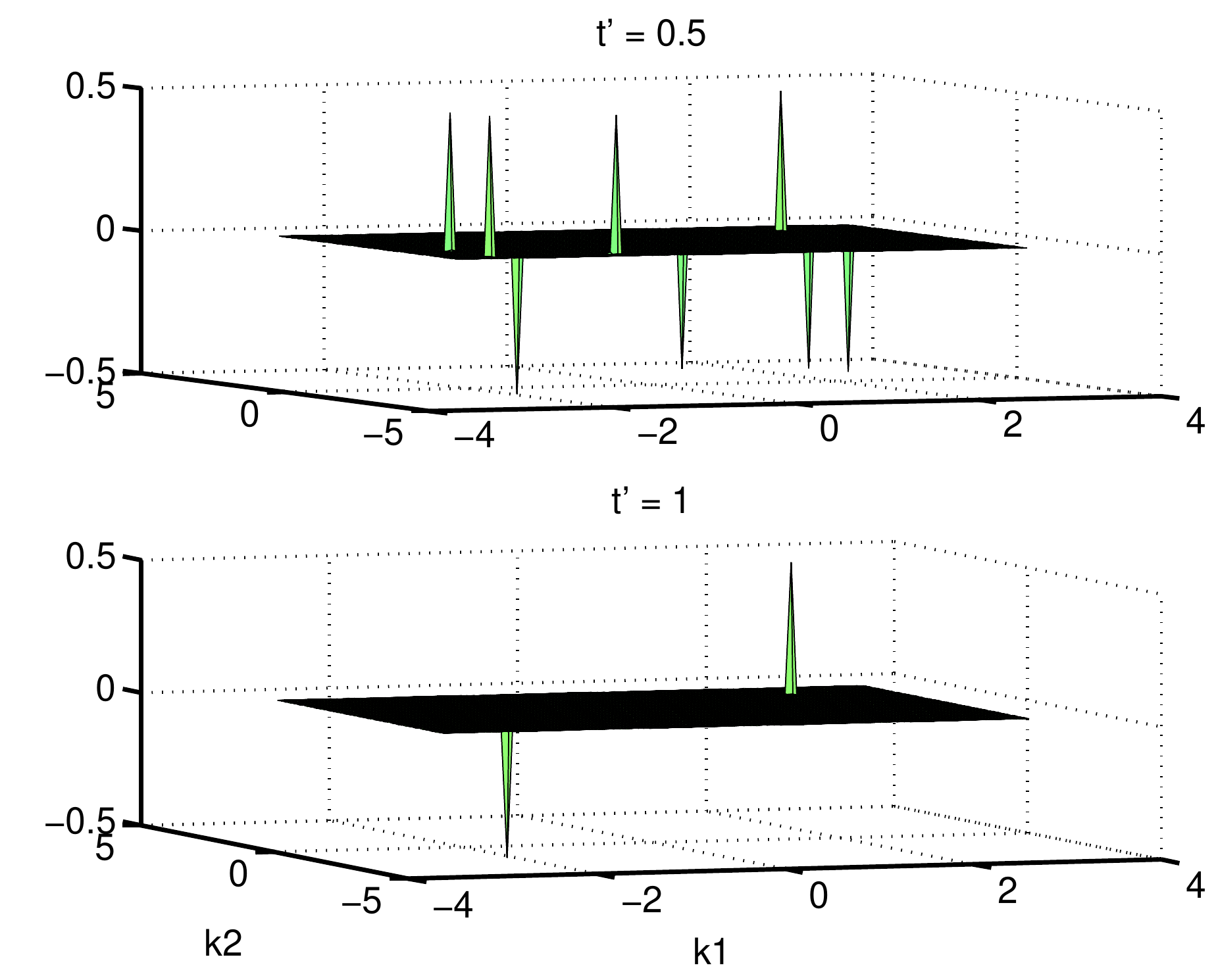}
 \caption{(color online) PB Curvature as a function of $k_1$ and $k_2$ for two different $t'$ values for the second band. The peaks correspond to the location of the DPs. As we change the $t'$ the number of DPs in the system changes.}.
 \label{fig:berry_curvature}
 \end{center}
 \end{figure}

\begin{figure}[ht]
 \begin{center}
 \includegraphics[width=7cm]{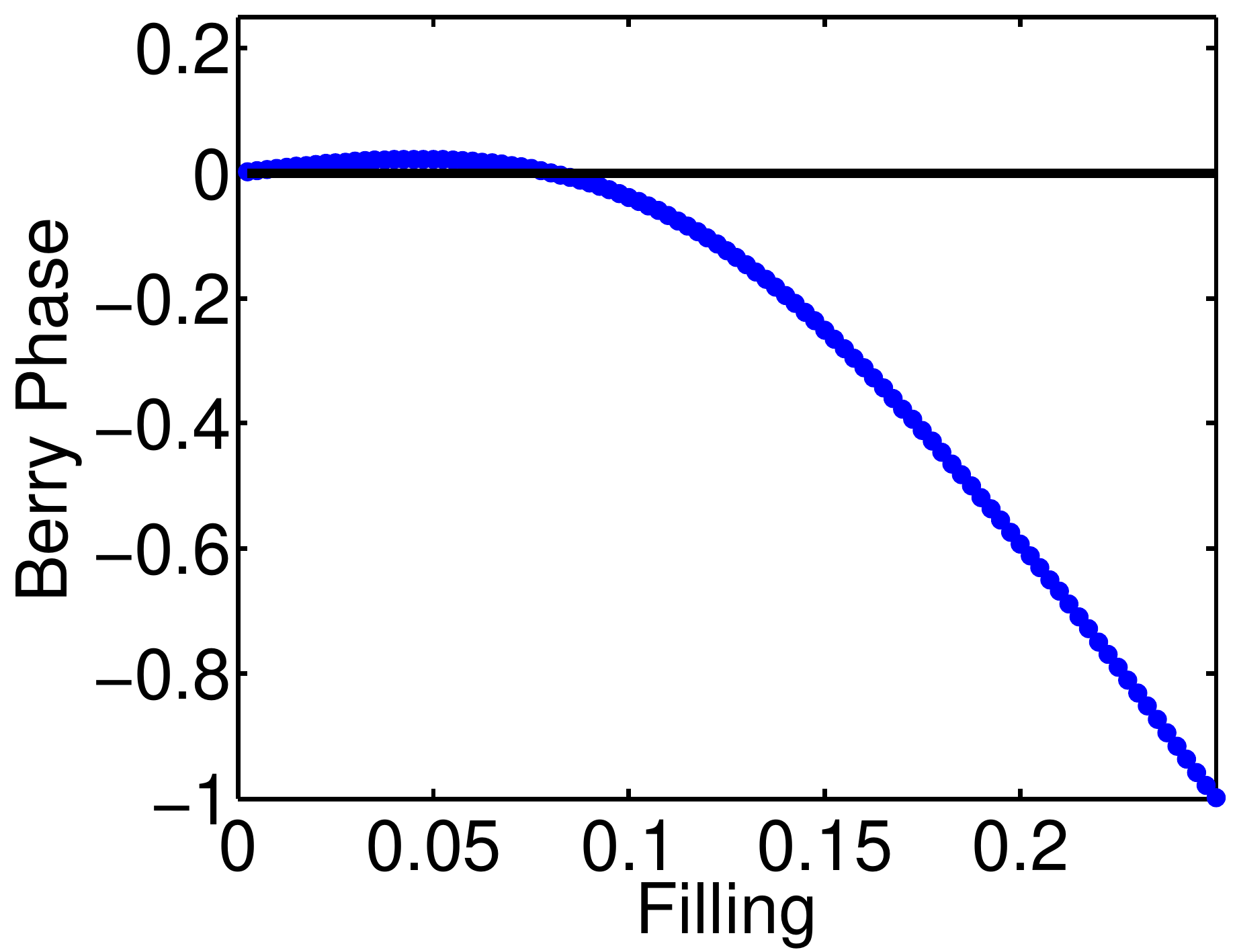}
 \caption{(color online) PB phase as a function of filling for $t' = 1$. The PB phase is not negative for all values of the filling. For some values it is positive reflecting that the PB curvature of the band takes both positive and negative values.}
 \label{fig:berry_phase_filling}
 \end{center}
 \end{figure}

At the DPs, as shown earlier the PB curvature ${\mathcal{B}}$ is singular. Applying a small staggered mass term to induce a gap at the DPs we compute the PB curvature of the second band at two $t'$ values, shown in FIG.\eqref{fig:berry_curvature}. The PB curvature of the second band shows a peak at the DPs. 
%
Using the PB curvature $\mathcal{B}$ obtain the Chern numbers, $\nu_n = -1$ for $n=1,4$ and $\nu_n = 1$ for $n=2,3$. At half-filling, the total Chern number given by $\nu = \nu_1 + \nu_2$ vanishes, implying that the Hall conductance also vanishes \cite{hassan2013, hassan2013b}. Remarkably, even though the Chern number for the lowest and the highest bands are both equal to $-1$, the ${\mathcal{B}}^n(k)$ for these bands is not negative for all values of $k$. This surprising result is true for the other bands as well with the signs flipped appropriately. FIG.\eqref{fig:berry_phase_filling} shows the PB phase as a function of filling for the lowest band, clearly depicting this behaviour. A consequence of this behaviour, can be seen in optical lattice experiments which shall be further discussed in section (\ref{sec:curv_kh_model}). Thus the non-interacting KHUB with PBC shows intriguing topological character.
%

\subsection{Topology of KHUB with OBC}\label{sec:edge_states}
The properties of the KHUB that were discussed till now are for PBC. We study the edge states in this model in a cylindrical geometry with zig-zag edges. There are zero energy edge states between the second and the third band and chiral edge states between the bottom two and the upper two bands. The number and the location of the zero-energy edge states in the quasi-momentum direction $k$ change as a function of $t'$ which can be determined using the Zak phase \cite{zak1989, saket2013, delplace2011} around a closed contour. There have been proposals to probe these phases in optical lattices \cite{atala2012}.  

\begin{figure}[ht]
 \begin{center}
 \includegraphics[width=7.5cm]{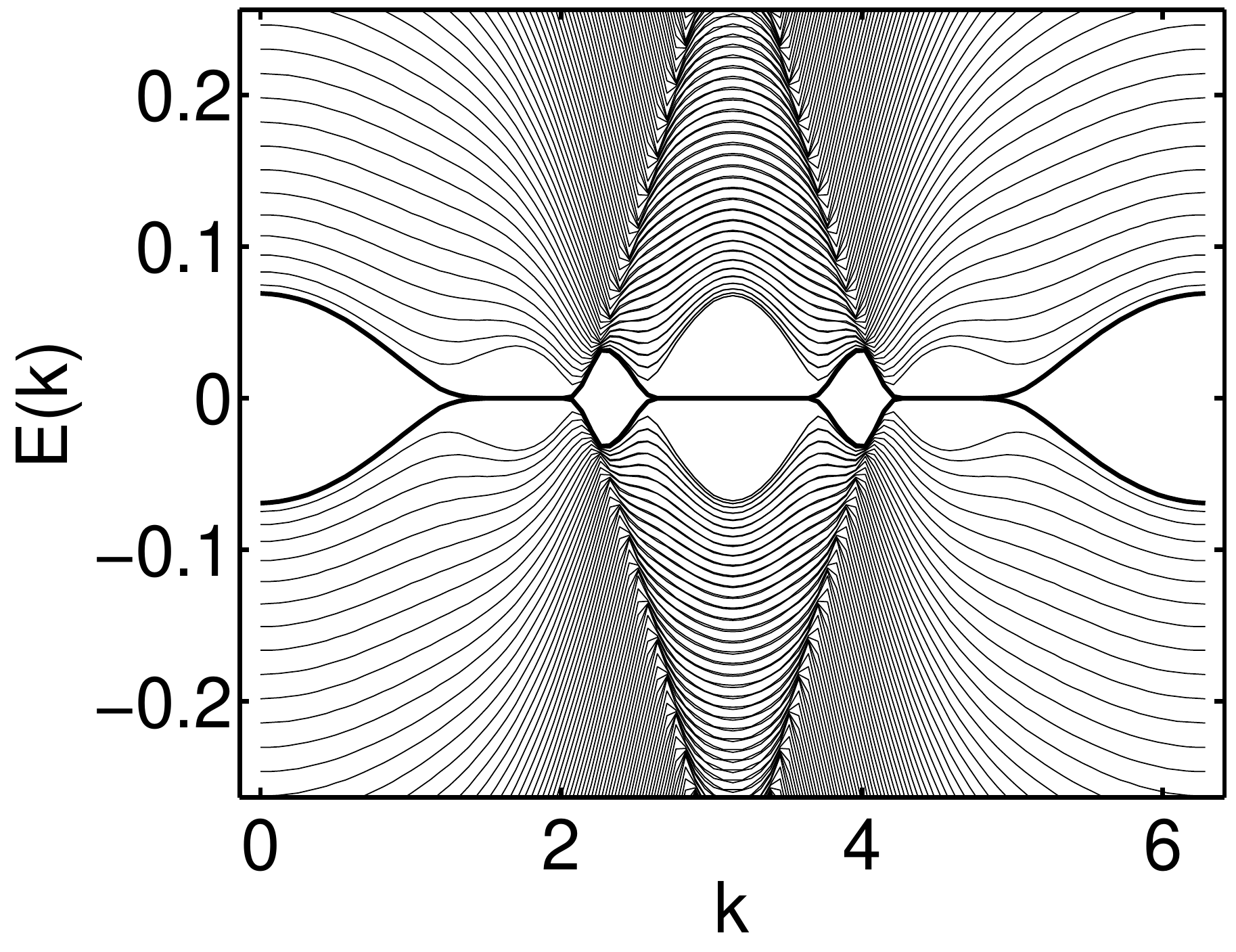}
 \includegraphics[trim = -25mm 0mm 15mm 0mm,width=6.7cm]{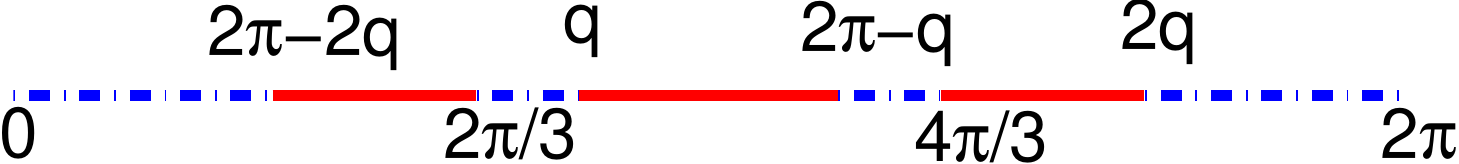}
\caption{(color online) The figure on the top panel is the zoomed spectrum for the energy of the KHUB with OBC showing the zero energy edge states. On the red solid lines of the figure in the bottom panel, the Zak phase is $+1$ whose correspondence to the existence of the edge states in the figure on the top panel can be seen.}
\label{fig:edgeenergy_bw}
 \end{center}
 \end{figure}

At the graphene limit, $t' = 0$, there are $2\pi/3$ continuous zero-energy edge states for each of the two spin species for $k \in (2\pi/3, 4\pi/3)$, making a total of $4\pi/3$ states. Here the Zak phase is $+1$ for $k \in (2\pi/3, 4\pi/3)$, and $0$ elsewhere. For values of $0<t'<1/\sqrt{3}$ between these two limits the edge states are not continuous, Fig:\eqref{fig:edgeenergy_bw}. The doubly-degenerate edge states in $k \in (2 \pi / 3, q) \,\cup\, (2 \pi - q, 4 \pi / 3)$ shift to $k \in (2\pi-2k, 2\pi/3) \,\cup\, (4\pi/3, 2k)$, respectively, forming unique states and thus preserving the total number. On the other hand, for $1/\sqrt{3} < t' <\sqrt{3}$, there are unique continuous edge states for $k \in (-2\pi/3 , 2\pi/3)$. Beyond $t' = \sqrt{3}$ again patches of zero-energy edge states occur. 

\begin{figure}[ht]
 \begin{center}
 \includegraphics[width=7cm]{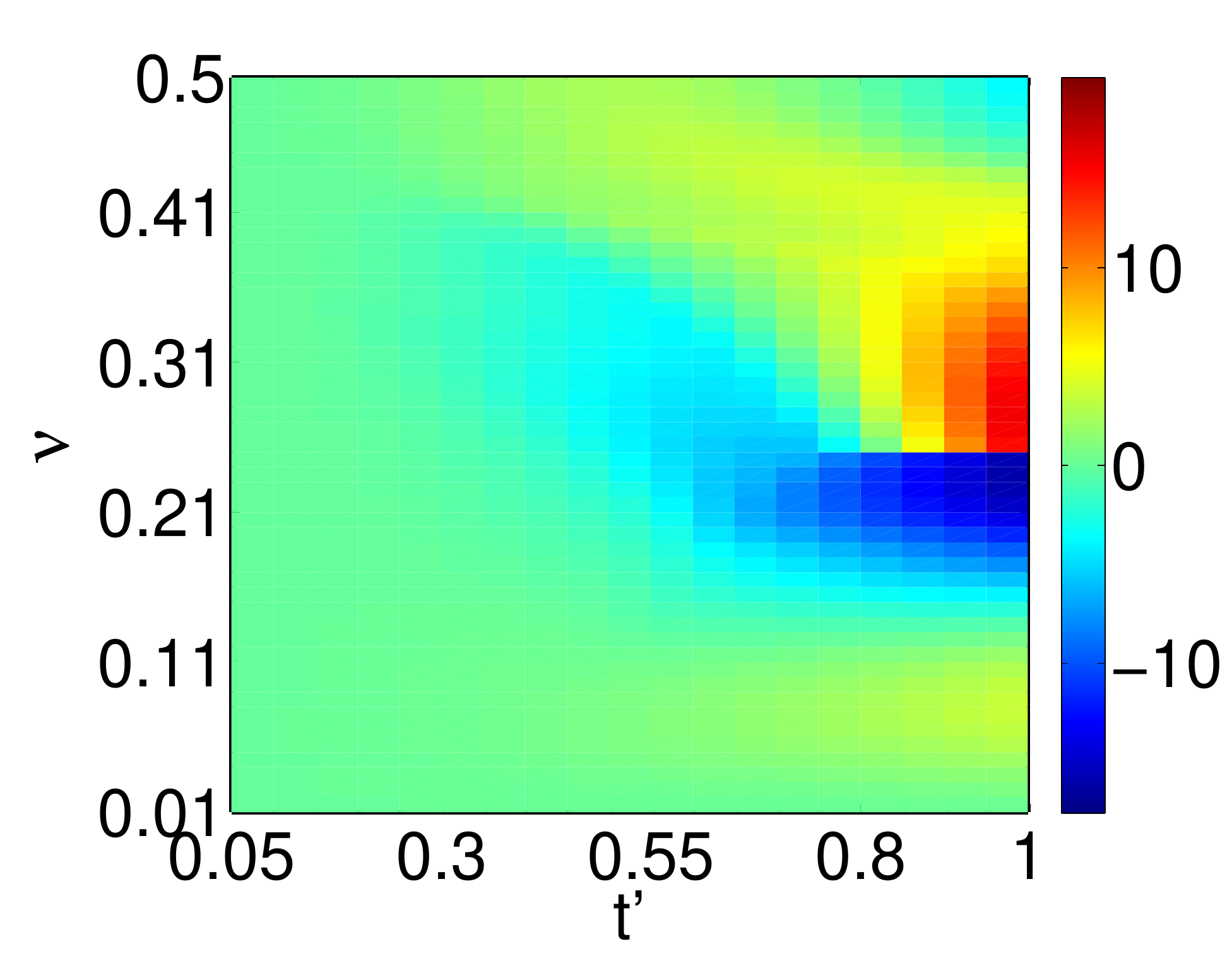}
 \caption{(color online) The total average charge current for an open tube of circumference $L = 140$ with zig-zag edges for various $t'$. Note that the sign of the edge current changes as a function for filling beyond the merging of the DPs at $t' = 1/\sqrt{3}$.}
 \label{fig:fulledgecurrent}
 \end{center}
 \end{figure}

The edge states carry current due to the breaking of time-reversal symmetry in the model. Using the Heisenberg equation of motion for the density operator and the density-current continuity equation, we can compute a general expression for the charge current between two sites on each of the $X$, $Y$ and $Z$ links. From the current along the $Z$ link 
\begin{align}
J^Z(i_1,i_2;i_1,i_2) & = i\sum_{\mu\sigma} a^\dagger_{i_1,i_2,\sigma} P^z_{\sigma,\mu} b_{i_1,i_2,\mu}  - h.c.,
\end{align}
the average charge current for cylindrical geometry can be computed
\begin{align}
J^z(i_1;i_1) & = \sum_{i_2}  J^z(i_1,i_2;i_1,i_2) \\
& = i \sum_{k_2} \sum_{\mu\sigma} a^\dagger_{i_1,k_2,\sigma} P^z_{\sigma,\mu} b_{i_1,k_2,\mu}  - h.c.
\end{align}
Similarly the average current on the $y$ link can also be calculated. 
 \begin{figure}[ht]
 \begin{center}
 \includegraphics[width=4.2cm]{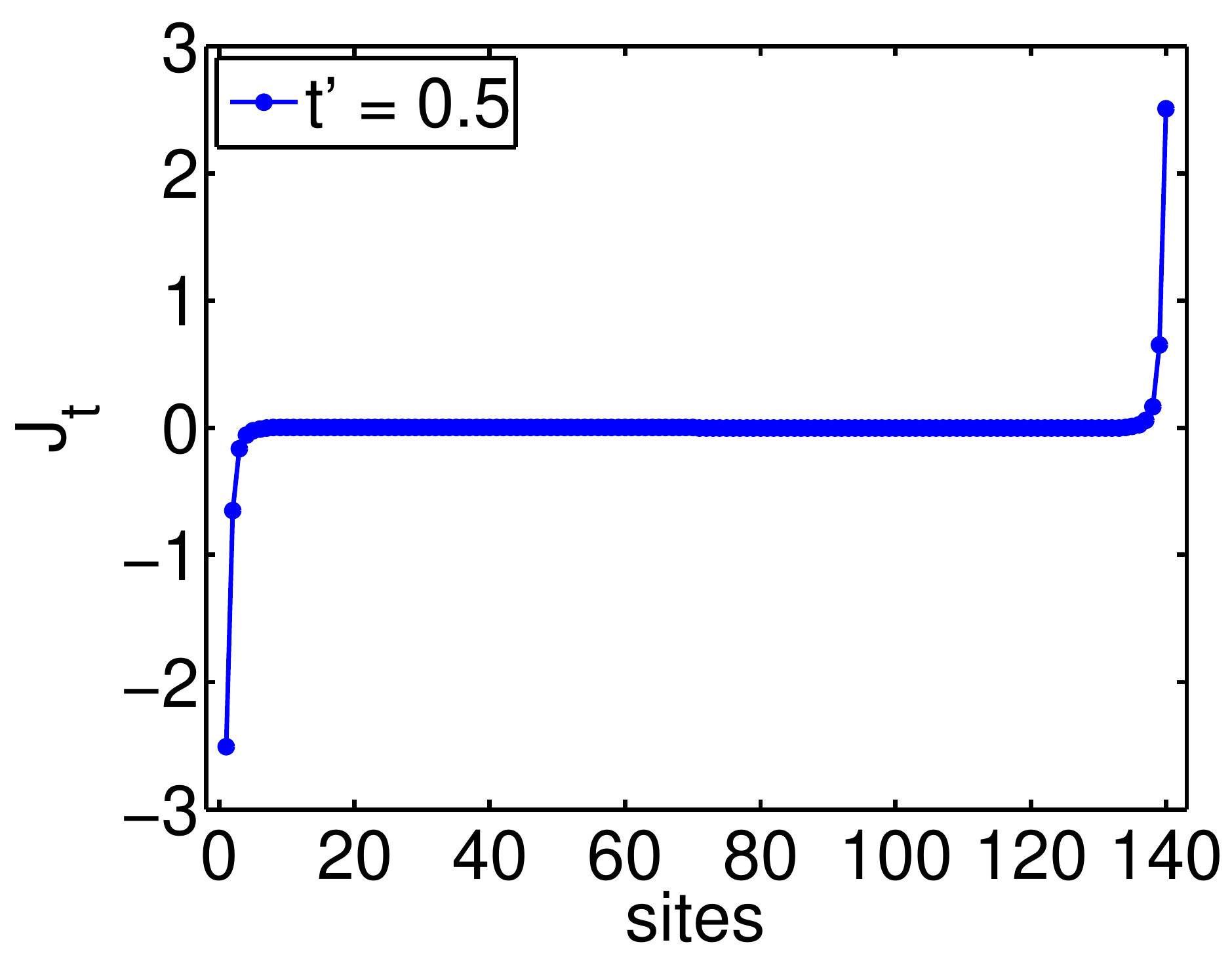}
 \includegraphics[width=4.2cm]{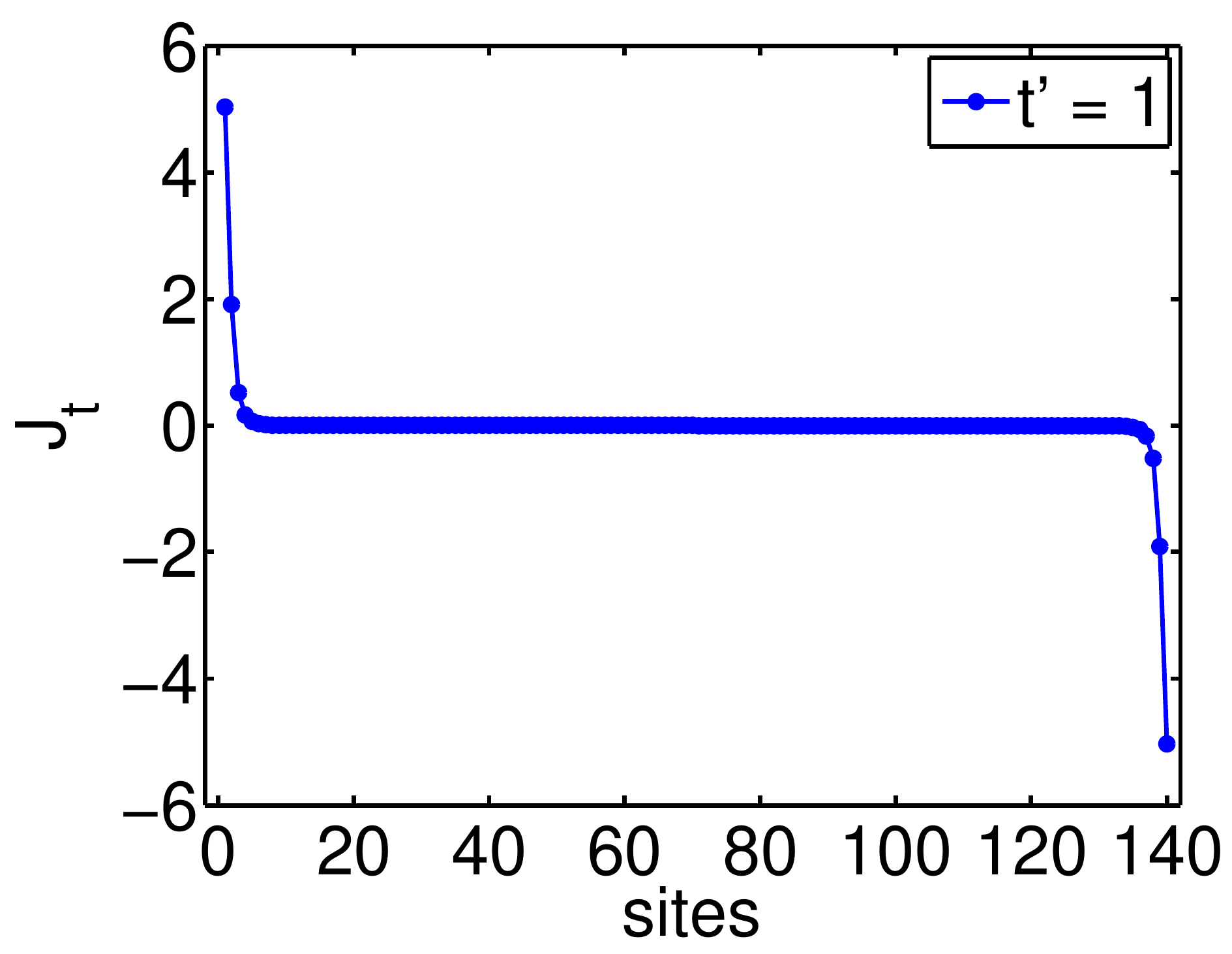}
 \caption{(color online) The total average charge current at $t' = 0.5t$ and $t' = t$. The current at the edges changes sign as a function of $t'$.}
 \label{fig:total_charge_tp_full}
 \end{center}
 \end{figure}
%
Using these expressions, we find that the total average charge current explicitly involves the time-reversal breaking spin-dependent strength $t'$. Thus the existence of non-zero currents at the two edges of the system in FIG.\eqref{fig:total_charge_tp_full} can be attributed to a non-zero $t'$. At $t'=0$ there is no current at the edges, as expected. As $t'>0$ a non-zero edge current appears and is initially negative at the left edge and positive at the right edge. However, by $t'=1$ the signs of the currents on the two edges have flipped. The sign of the edge current also depends on which states are filled for $t' > 0.717$. Since the chiral edge states have an opposite and larger contribution than those in the bulk, the edge current flips sign at quarter filling when the former begins to fill, as shown in FIG.\eqref{fig:fulledgecurrent}.

These rich topological properties of the model motivates the study of realizing the model and the properties in optical lattice experiments.

\section{Experimental Realization}
We study the realization of the model and the above mentioned topological features in cold atom experiments in the following sections. The model can be realized in optical lattice systems as shown by Duan {\em et al.}\cite{duan2003}. We systematically derive the hamiltonian in the next section. We then discuss methods of probing the DPs using Bloch-Zener oscillations and we propose a method to detect the PB curvature in optical lattice systems. 
\subsection{Spin-dependent hopping in the honeycomb lattice}
The KHUB is a model on the honeycomb lattice with spin-dependent hopping. There are multiple methods of obtaining spin-independent hopping on the honeycomb lattice \cite{loon2010}. For example, the required honeycomb lattice can be generated by three intersecting laser beams at an angle of $120^{\circ}$ between them \cite{duan2003, loon2010}. In this section, we systematically derive the spin-dependent hopping on the honeycomb lattice using the method suggested by Duan {\em et. al.} \cite{duan2003} which is different from that discussed earlier \cite{mazza2012}.

\begin{figure}[ht]
 \begin{center}
 \includegraphics[width=4cm]{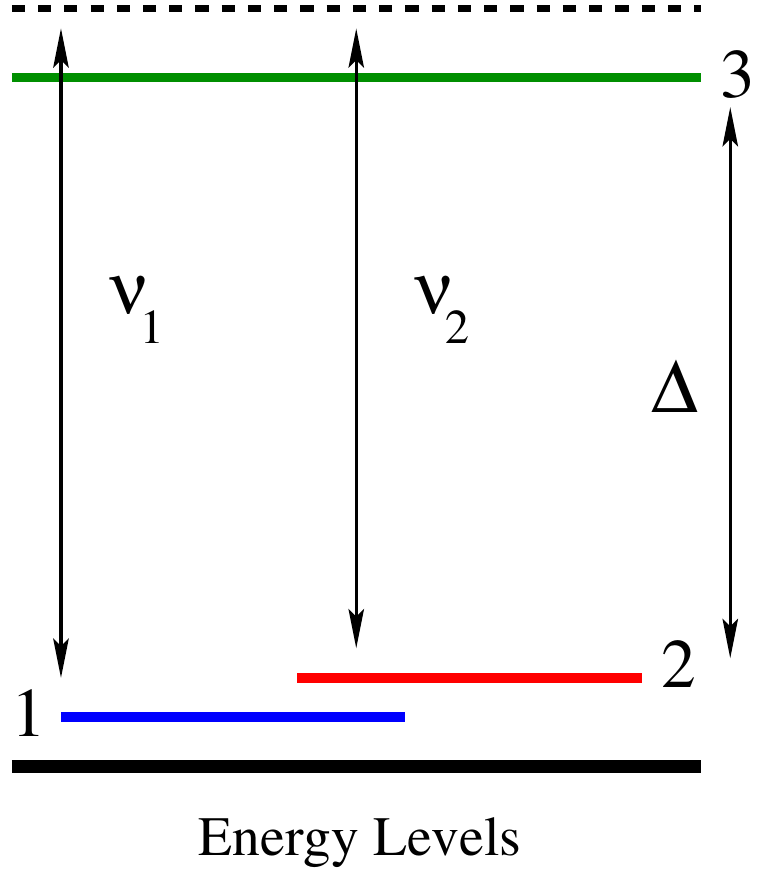}
 \caption{(color online) Schematic of the effective hyperfine energy levels of $^{40}K$. The gap $\Delta$ is orders of magnitude larger in energy compared to the lower energy levels \cite{leblanc2006}.}
 \label{fig:energy_levels}
 \end{center}
 \end{figure}

Most fermionic optical lattice experiments are performed using $^{40}K$ atoms \cite{leblanc2006}. In the absence of external magnetic field the $^2S_{1/2}$ and the $^2P_{1/2}$ levels of potassium each split into two hyperfine levels. Two of the hyperfine energy levels of $^2S_{1/2}$ are much lower in energy compared to levels of $^2P_{1/2}$.  FIG. (\ref{fig:energy_levels}) is a schematic of the three level system formed by the low levels of $^2S_{1/2}$ and a level of $^2P_{1/2}$.
 
The lower two energy levels $1,2$ are separated from the third $3$ by a gap $\Delta \approx 0.3 eV$  which is orders of magnitude larger\cite{leblanc2006} than the hopping parameter $\approx 10^{-13} eV$ seen in typical optical lattice experiments  \cite{mazza2012}. Two blue de-tuned laser beams $L_1$ and $L_2$ excite virtual transitions between the first and the third ($1\rightarrow 3$) and the second and the third ($2\rightarrow 3$) levels respectively. Since these virtual transitions are fast compared to the hopping of the atoms, a local microscopic hamiltonian can be written as
\begin{align}
\mathcal{H} & = \int d^2x \sum_{i = 1}^{2} \epsilon_i C^\dagger_{i}({\bf{x}}) C_i({\bf{x}}) + \Delta C^\dagger_{3}({\bf{x}}) C_3({\bf{x}}) \nonumber\\
& ~~~~~+\sum_{i=1}^{2}g_i C_3^\dagger({\bf{x}}) a_{i}({\bf{x}},\tau) C_i({\bf{x}}) + h.c.
\end{align}
Here $\epsilon_i$ represents the energies of $1$ and $2$. $C_i$ is the atomic creation operator for the $i$-th energy level at ${\bf{x}}$. The last two terms in the above expression arise due to the interaction of the atom with the laser beams. The wavelengths of the laser beams is such that it only causes transitions from the energy levels $1$ and $2$ to the energy level $3$ with $g_1$ and $g_2$ representing the strength of these transitions respectively and $a_i({\bf{x}},\tau)$ represents the electromagnetic field. The effective two-level system can be obtained by integrating out the third energy level. The action $S$ in the path integral formalism for the three level system can be written as
\begin{align}
S & = -\int d\tau\left(\sum_{i=1}^3 C_i^\dagger({\bf x}) \partial_\tau C_i({\bf x}) + \mathcal{H}\right).
\end{align}
Integrating the third energy level we obtain the effective action of the two-level system as
\begin{align}
S_e & = \int d\tau d\tau' \int d^2x \sum_{i,j = 1}^{2} C^\dagger_{i}({\bf x},\tau) G_{ij}({\bf x},\tau,\tau') C_j({\bf x},\tau') 
\end{align}
where the matrix $G$ is given as
\begin{align}
G_{ij}({\bf x},\tau,\tau') & = \partial_\tau \delta_{ij}+ \epsilon_i({\bf x})\delta_{ij} \nonumber\\
& ~~~+ g_ig_j a_i^{*}({\bf x},\tau)\langle \tau | \frac{1}{\partial_\tau+\Delta} |\tau'\rangle  a_j({\bf x},\tau')  \label{greenfun}
\end{align}
Since the beams are monochromatic, the electromagnetic fields can be written as $a_i({\bf x},\tau) = e^{i\nu_i\tau}b({\bf x})$, where $\nu_i$ is the frequency of transition from the $i$-th, $i = 1,2$, energy level to the third energy level. This is clearly seen in FIG.\eqref{fig:energy_levels}. Thus the effective new $G $ matrix is given by 
\begin{align}
G_{ij}({\bf x},\tau,\tau') & =  \partial_\tau \delta_{ij}+ \epsilon_i({\bf x})\delta_{ij} +  \frac{g_ig_j}{\Delta} b_i^{*}({\bf x})b_j({\bf x}). 
\label{effgreenfun}
\end{align}
The two low-lying energy levels labelled $1$ and $2$ can be represented by pseudo-spin indices $\sigma = -1$ and $\sigma = +1$ respectively. The effective potential seen by the pseudo-spins is given as
\begin{align}
V_s^{\sigma\sigma'}({\bf x}) & = g_\sigma g_{\sigma'} b_{\sigma}^{*}({\bf x})b_{\sigma'}({\bf x}).
\end{align}
When $N$ laser beams of varying intensities and directed along the wave-vectors ${\bf k}_n$ are incident on this effective two-level atom, the field can be written as
\begin{align}
b_{\sigma}^{*} &= \sum_n a_{n\sigma} \sin({\bf k}_n \cdot {\bf x})
\end{align}
with the condition $\langle a_{n\sigma}^* a_{n'\sigma'} \rangle = \delta_{nn'} \langle a_{n\sigma}^* a_{n\sigma'} \rangle$. This implies that the fields arising due to different laser beams are independent. The effective potential which depends on the spin becomes 
\begin{align}
V_s^{\sigma\sigma'}({\bf x}) & = g_\sigma g_{\sigma'} \sum_n \sin^2 ({\bf k}_n \cdot {\bf x}) a_{n\sigma}^{'*}a'_{n\sigma'}.
\end{align}
Thus the spin dependent potential can be varied by tuning the laser beams $L_1$ and $L_2$. 

The spin-dependent potential on the honeycomb lattice can be generated by tuning three lasers with different strengths oriented along the three directions $X$, $Y$ and $Z$, at an angle of $120^{\circ}$ relative to one another \cite{duan2003, loon2010}, FIG.\eqref{fig:honeycomb}. The $L_1$ laser beam is sufficient to generate a spin dependent coupling on the $Z$ link, $a_{Z\downarrow} = 0$. The lasers $L_1$ and $L_2$ directed along $X$ and $Y$ with a relative phase difference are required to generate the couplings along these directions. Thus we have $g_{\uparrow} a_{X\uparrow} = g_{\downarrow}a_{X\downarrow}$ and $g_{\uparrow} a_{Y\uparrow} = ig_{\downarrow}a_{Y\downarrow}$ respectively for the $X$ and $Y$ directions. We now write the potential as a sum of spin-independent and spin-dependent parts, 
\begin{align}
V_{s}({\bf x}) & = V({\bf x})\mathbb{I} + {\bf{B}}({\bf x})\cdot\boldsymbol{\sigma}
\end{align}
where ${\bf{B}}({\bf x})$ is the effective space dependent magnetic field generated by the laser beams. Its individual components can be written as 
\begin{align}
B_x({\bf x}) & = g_{\uparrow}^2 |a'_{x\uparrow}|^2 \sin^2 ({\bf k}_X \cdot {\bf x}) \\
B_y({\bf x}) & = g_{\uparrow}^2 |a'_{y\uparrow}|^2 \sin^2 ({\bf k}_Y \cdot {\bf x}) \\
B_z({\bf x}) & = \frac{1}{2}g_{\uparrow}^2 |a'_{z\uparrow}|^2 \sin^2 ({\bf k}_Z \cdot {\bf x})
\end{align}
where ${\bf k}_{X,Y,Z}$ is the wave-vector along $X,Y$ and $Z$ respectively. The spin independent potential $V({\bf x})$ can be written as
\begin{align}
V({\bf x}) & = g_{\uparrow}^2 |a'_{x\uparrow}|^2 \sin^2 ({\bf k}_X \cdot {\bf x}) + g_{\uparrow}^2 |a'_{y\uparrow}|^2 \sin^2 ({\bf k}_Y \cdot {\bf x}) \nonumber\\
& ~~~+ \frac{1}{2}g_{\uparrow}^2 |a'_{z\uparrow}|^2 \sin^2 ({\bf k}_Z \cdot {\bf x})
\end{align}
This shows that the spin-independent part $V$  cannot be tuned individually as it is coupled to ${\bf B}({\bf x})$. So we add an additional spin-independent potential $V_h^{\sigma\sigma'}({\bf x})$ which can be tuned without affecting the spin-dependent part. Now the wave-function of the atom in the spin-dependent honeycomb lattice potential follows the time-independent Schr\"{o}dinger equation of the form
\begin{align}
\left(\frac{p^2}{2M}\delta_{\sigma\sigma'} + V^{\sigma\sigma'}({\bf x})\right)\psi_{\sigma'}({\bf x}) & = E \psi_{\sigma}({\bf x})
\end{align}
where $V^{\sigma\sigma'}({\bf x}) = V_h^{\sigma\sigma'}({\bf x}) + V_s^{\sigma\sigma'}({\bf x})$ is the total potential. The wave-function of the atoms can be expanded in terms of atomic orbitals $\phi_\alpha({\bf x} - {\bf x}_i)$ which are localized in the $\alpha^{th}$ sublattice of the $i$-th triangular Bravais lattice, that is
\begin{align}
\psi_{\sigma'}({\bf x}) & = \sum_{i\alpha} d_{i\alpha}^{\sigma'} \phi_\alpha^{\sigma'}({\bf x} - {\bf x}_i).
\end{align} 
This allows us to compute the hopping parameter as
\begin{align}
t_{i\alpha j\beta}^{\sigma'\sigma} & = \int d^2x~\phi_\beta^{*\sigma}({\bf x} - {\bf x}_j) \left(\frac{p^2}{2M} + V^{\sigma\sigma'}({\bf x})\right) \phi_\alpha^{\sigma'}({\bf x} - {\bf x}_i).
\end{align}
Eq.\ (\ref{genham}), with the hamiltonian Eq.\eqref{khk}, is applicable when the nearest neighbours provide the dominant contributions to the hopping. Thus we need $i = j \pm 1$ for the $X$ and $Y$ links and $i = j$ for the $Z$ link with $\alpha \neq \beta$. Once we obtain the spin-dependent hopping the onsite interactions between the atoms in the optical lattice systems can be created by Feshbach resonance\cite{loon2010}. At very large interactions both the Kitaev-Heisenberg model \cite{chaloupka2010,duan2003,hassan2013b,hassan2013} and the isotropic Kitaev model \cite{kitaev2006} can be obtained. Thus the anisotropic Kitaev model can be obtained by tuning the strengths of individual laser beams, and this model can be further mapped onto the toric-code hamiltonian \cite{kitaev2003}.


\subsection{Probing the topological properties of the model}
\subsubsection{Locating the DPs: Bloch oscillations}

Recently an experimental method to probe the DPs using Bloch-Zener oscillations \cite{breid2006,kolovsky2003} was suggested by Tarruell {\em et al.} \cite{tarruell2012}, whereas a detailed method of numerically simulating such oscillations was discussed by Uehlinger {\em et al.} \cite{uehlinger2013}. Here we probe the DPs in our model using Bloch-Zener oscillations. 

At time $T = 0$, the tight binding Kitaev-Hubbard hamiltonian with a staggered potential and in the presence of a harmonic trap is given by
\begin{align}
\mathcal{H}_0 & = -\sum_{<ij>}C_{i\mu}^{\dagger}\frac{(tI +t'\sigma^{\alpha})_{\mu \nu}}{2}C_{j\nu} + \frac{W}{2}\sum_{i\in A} n_{i} \nonumber\\
&~~~- \frac{W}{2}\sum_{i\in B} n_{i} + \sum_i (\gamma_x x_i^2 + \gamma_y y_i^2) n_i \label{opticalhami}
\end{align}
Here $W$ is the strength of the staggered onsite potential, $\gamma_x$ and $\gamma_y$ are the strengths of the harmonic trap in the $\hat{e}_1$ and $\hat{e}_2$ directions, while $x_i$ and $y_i$ represent the spatial coordinates of the $i^{th}$ lattice site which are measured in terms of the lattice parameter $a$. 

We calculate the $n$-particle many body ground state $|\psi(0)\rangle$ for this hamiltonian and  evolve it using the total hamiltonian $\mathcal{H} = \mathcal{H}_0 + \mathcal{H}_{int}$. Here the interaction term is that of an external force field of magnitude $F$ along $\hat{f}$ on the lattice, and is given by
\begin{align}
\mathcal{H}_{int} & = F \sum_i \hat{f}\cdot \hat{r}_i
\end{align}
where $\hat{r} = (x,y)$ is the position vector of the lattice site. The Schr\"{o}dinger evolution is
\begin{align}
|\psi(\tau) \rangle = e^{-i\mathcal{H}\tau} |\psi(0)\rangle
\end{align}
where $\tau$ is measured in terms of the Bloch oscillation time period $T_B = 2\pi/F$. We choose  $2\times 120^2$ lattice sites in order to prevent the cloud from ever hitting the boundary. At every time step we measure the projection of the Fourier-transformed many-body density matrix on to the density matrix of the single-particle bands in the presence of the staggered potential,  
\begin{align}
P_n(k_1,k_2,\tau) & = |\langle \chi_n|k_1,k_2\rangle \langle k_1,k_2 | \psi(\tau) \rangle|^2.
\end{align}
Here $|\chi_n\rangle$ is the single-particle eigen-state of the $n$-th band of the non-interacting hamiltonian with staggered mass. It is possible to project the density matrix because we have assumed that the trap potential varies slowly so that the single-particle bands do not change in the presence of the trap. 

We also compute probability amplitude per particle
\begin{align}
P_n(\tau) & = \frac{1}{N}\int \frac{d^{2}k}{4\pi^2} P_n(k_1,k_2,\tau)
\end{align}
where $N$ is the number of particles in the system. The quasi-momentum distribution of the particles clearly shows a sudden reduction in the density when a DP is encountered which gives us a method of probing them in experiments.

\begin{figure*}[ht]
 \begin{center}
 \includegraphics[width=4.4cm]{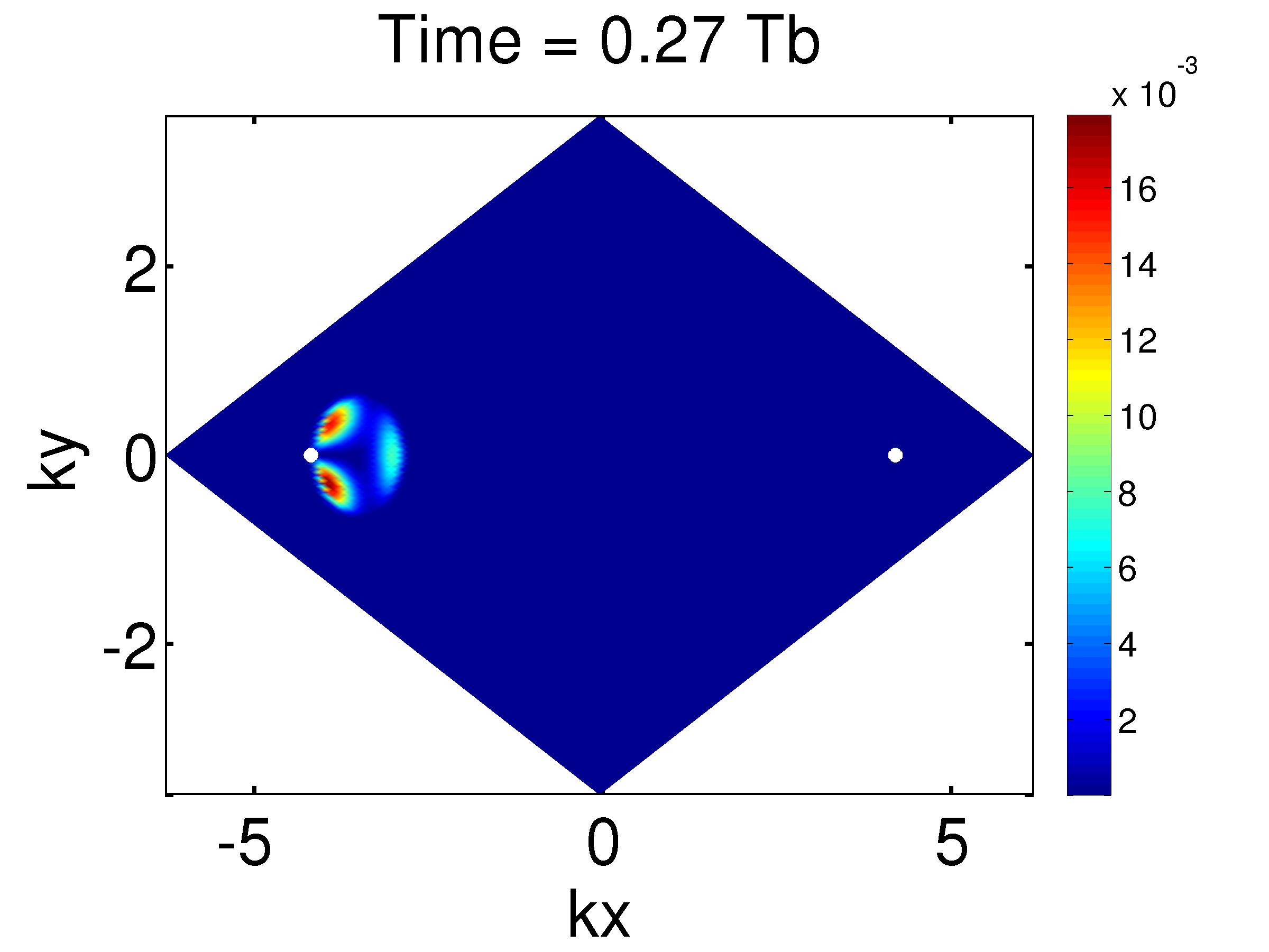}
 \includegraphics[width=4.4cm]{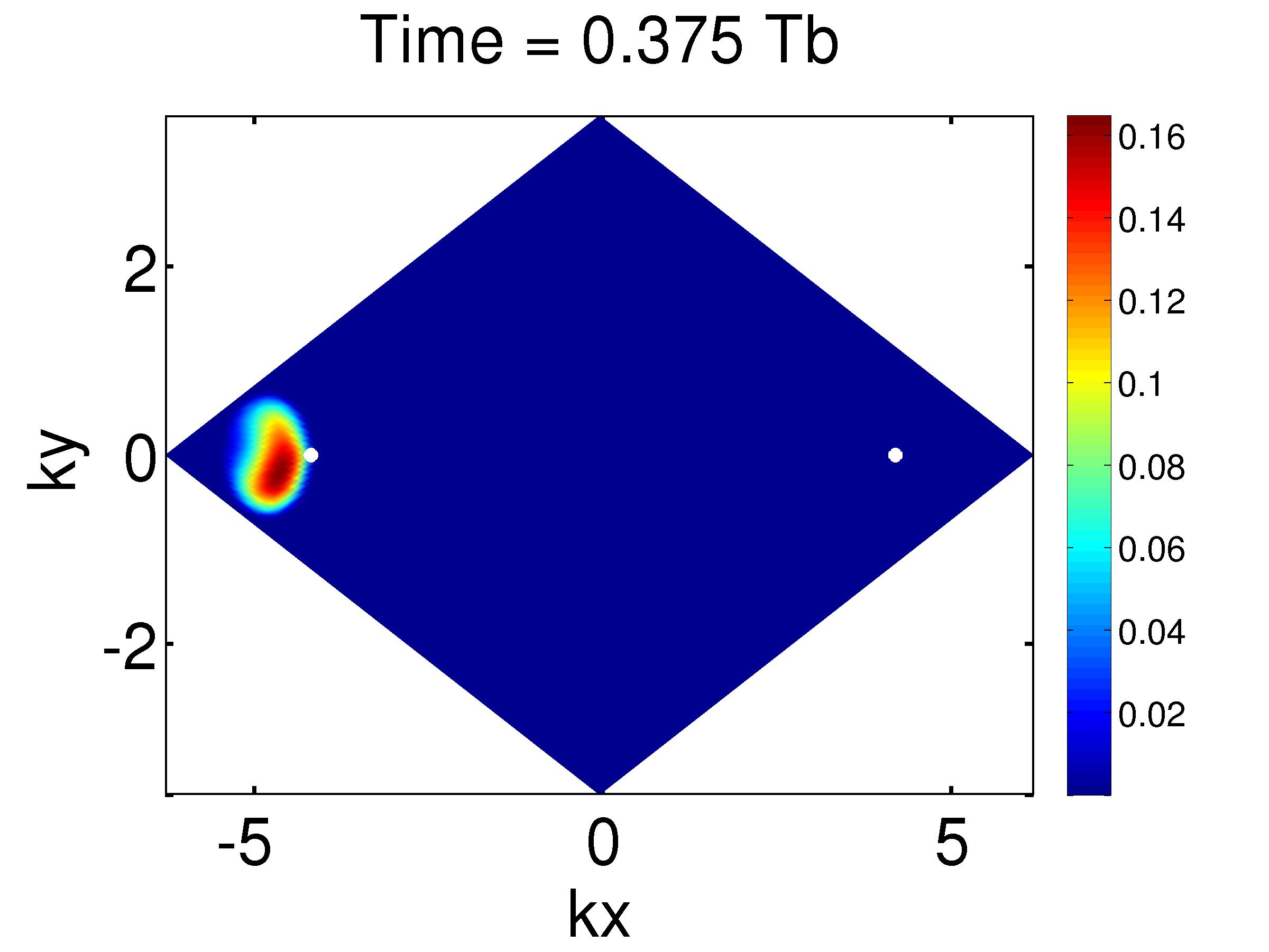}
 \includegraphics[width=4.4cm]{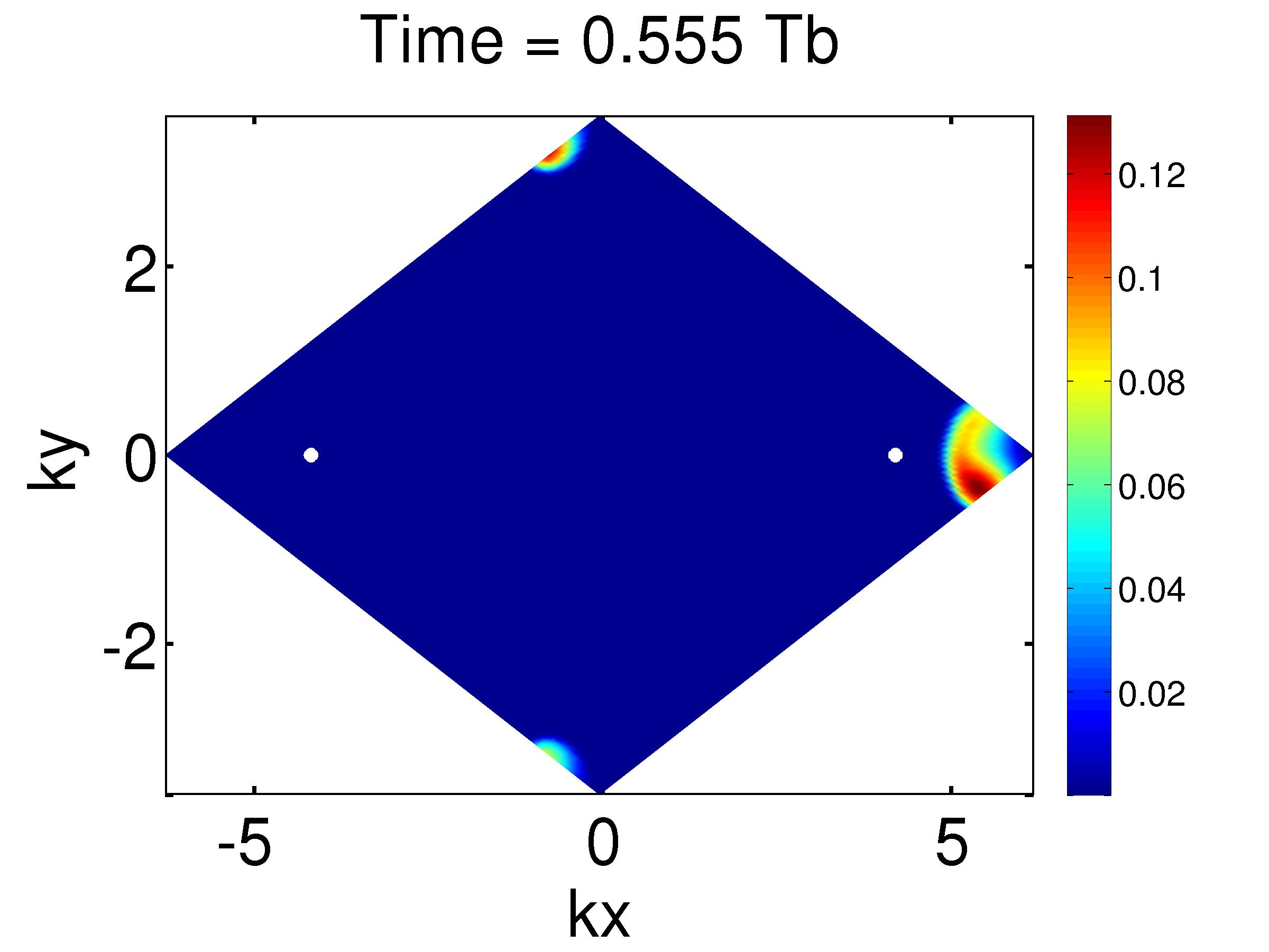}
 \includegraphics[width=4.4cm]{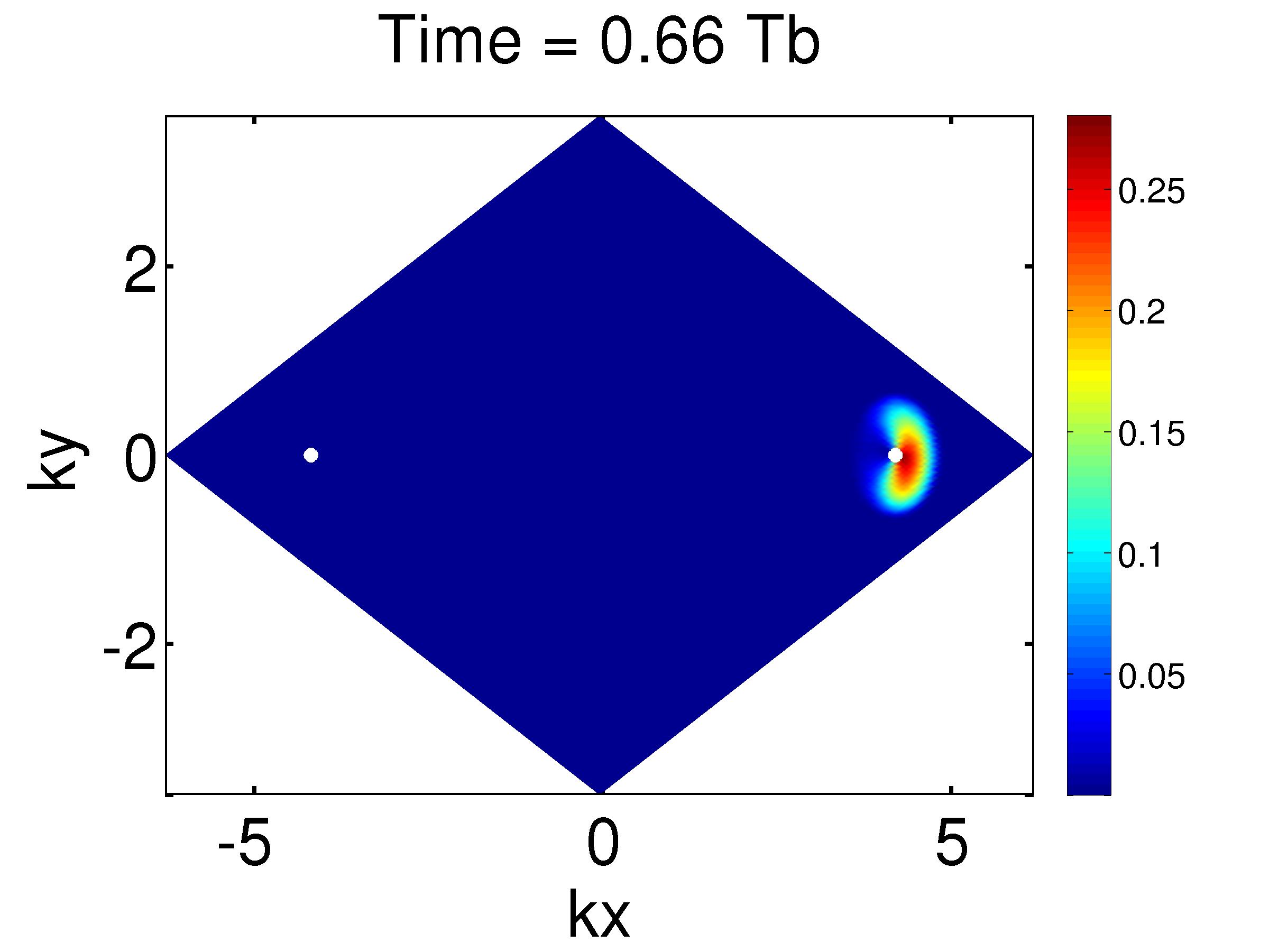}
 \caption{(color online) The quasi momentum distribution of the second band for Bloch-Zener oscillations as a function of $(k_x,k_y)$ at $t' = t$ resulting from a force acting along the $\hat{e}_1 + \hat{e}_2 = \hat{x}$ direction. Total number of particles considered is 187.}
 \label{fig:kh_100_xpyp}
 \end{center}
 \end{figure*}

We study the quasi-momentum distribution for $t'=t$ and $t'=0.5t$. In FIG.\eqref{fig:kh_100_xpyp} the quasi-momentum probability amplitude (in the orthogonal coordinates $(k_x,k_y)$) of $187$ particles in the second band for various instances during one Bloch oscillation is plotted. The parameters are $t/h = 589$Hz, $t' = t$, $F/h = 80$Hz, $\gamma_{x,y}/h = [0.01,0.01]$Hz, $W/h = 2$Hz and $\hat{f} = \hat{e}_1+\hat{e}_2$. The probability amplitude $P_n(k_x,k_y,\tau=0)$ initially localized around the origin, moves along the $k_x$ direction and encounters a DP at location $(-4\pi/3,0)$ at time $\tau = 0.27T_B$. 
%
%
On further evolution the DPs at $(4\pi/3,0)$ is probed finally returning to the center of the Brillouin zone after one oscillation. There is a transfer of particles to the higher bands close to the DPs as seen in FIG.\eqref{fig:occprob_xpyp} where we have plotted $P_l(\tau) = P_1(\tau) + P_2(\tau)$ and $P_u(\tau) =  P_3(\tau) + P_4(\tau)$. The two peaks in the figure corresponds to the DPs seen from the quasi-momentum distribution. The inset shows the probabilities for individual bands, with transitions occurring between each of the successive bands.

\begin{figure*}[ht]
 \begin{center}
 \includegraphics[width=4.4cm]{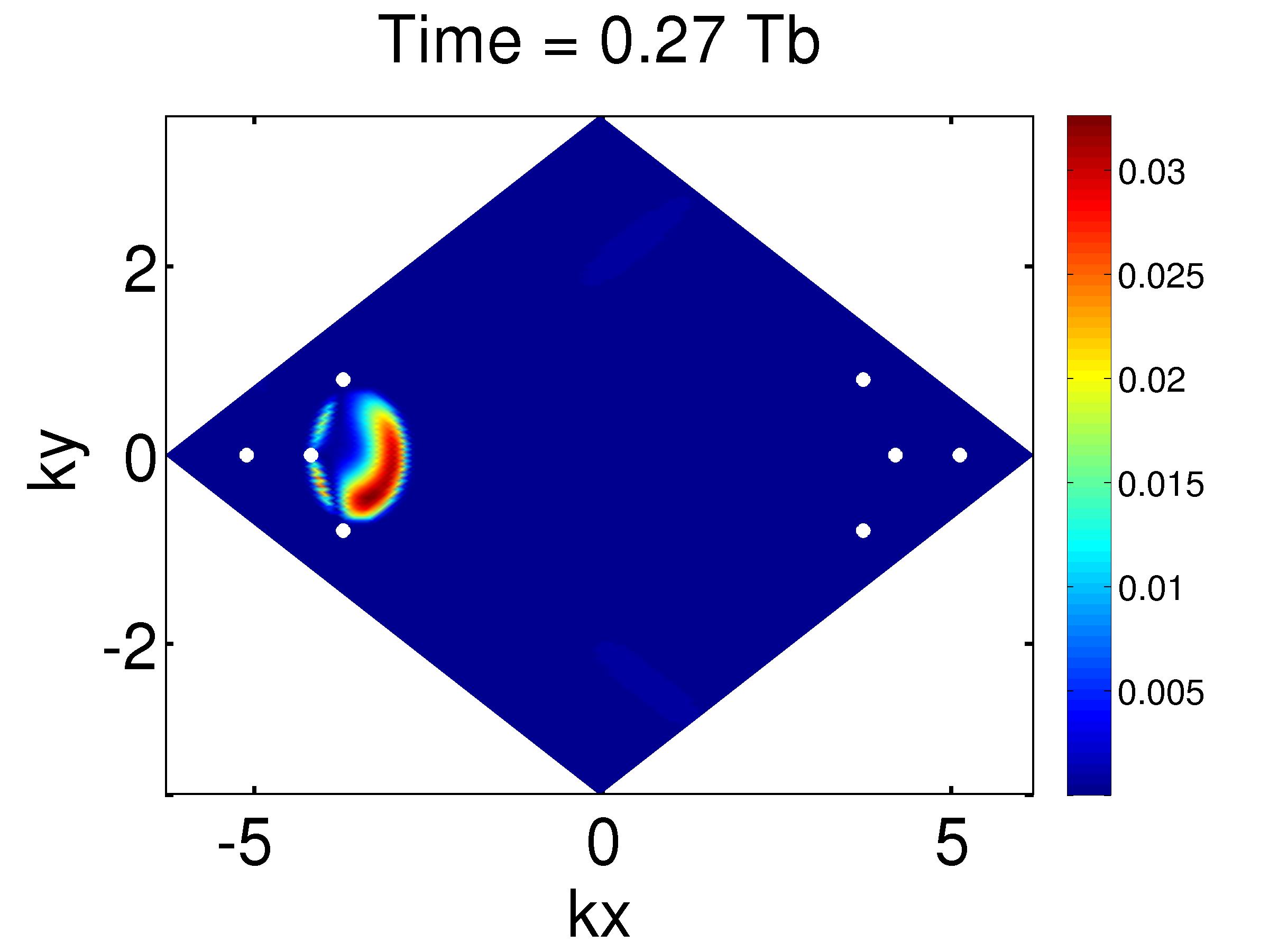}
 \includegraphics[width=4.4cm]{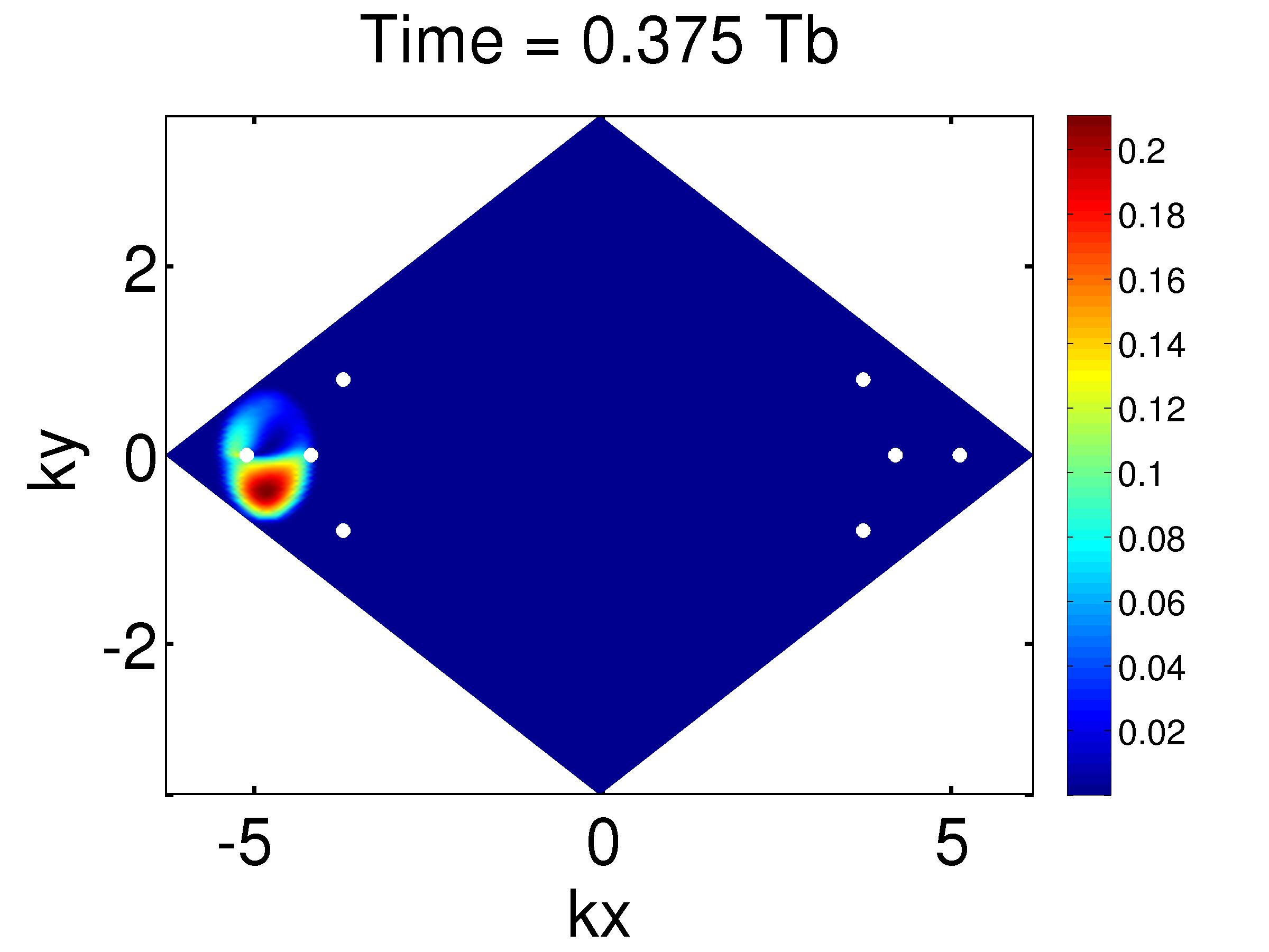}
 \includegraphics[width=4.4cm]{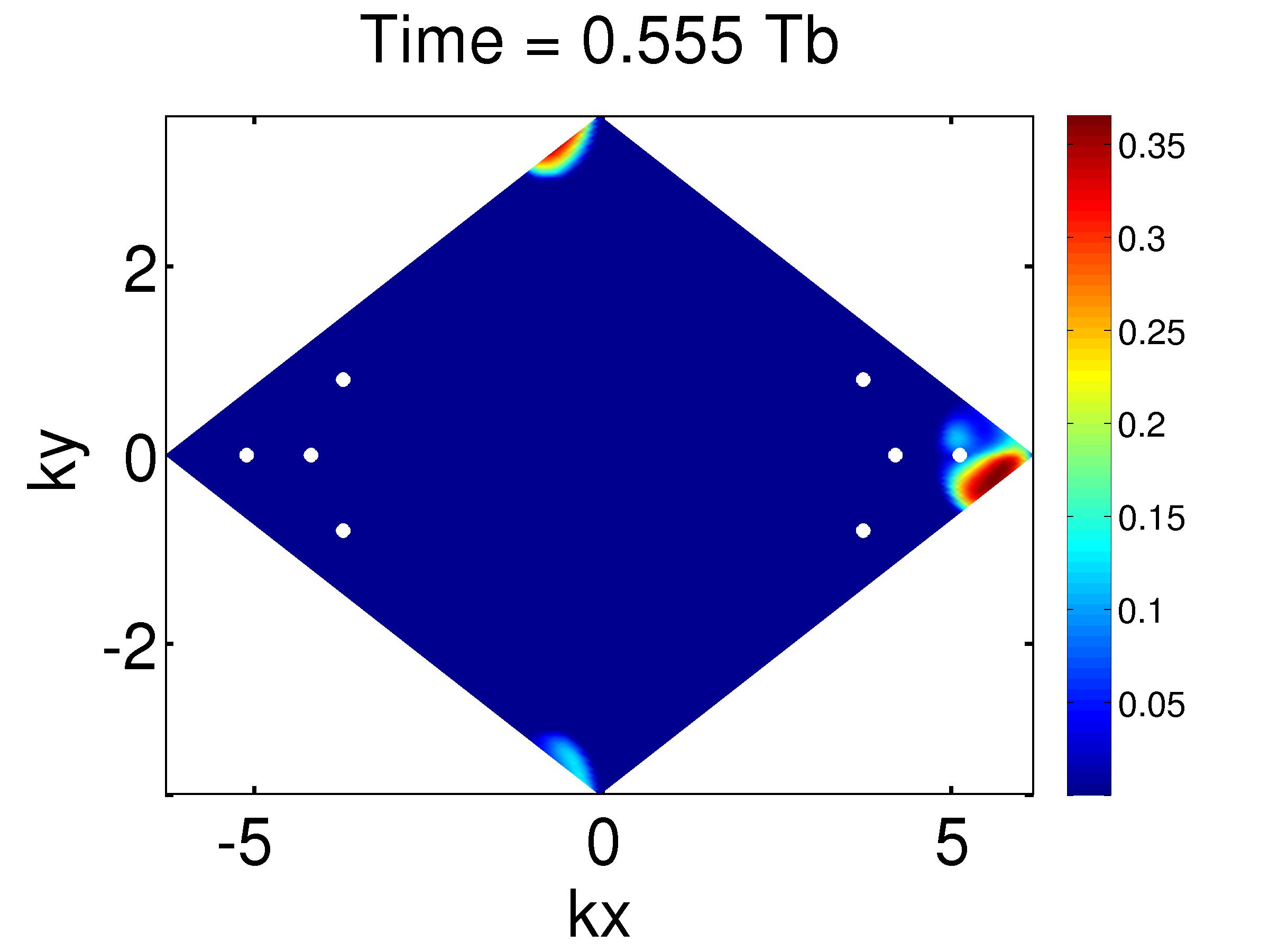}
 \includegraphics[width=4.4cm]{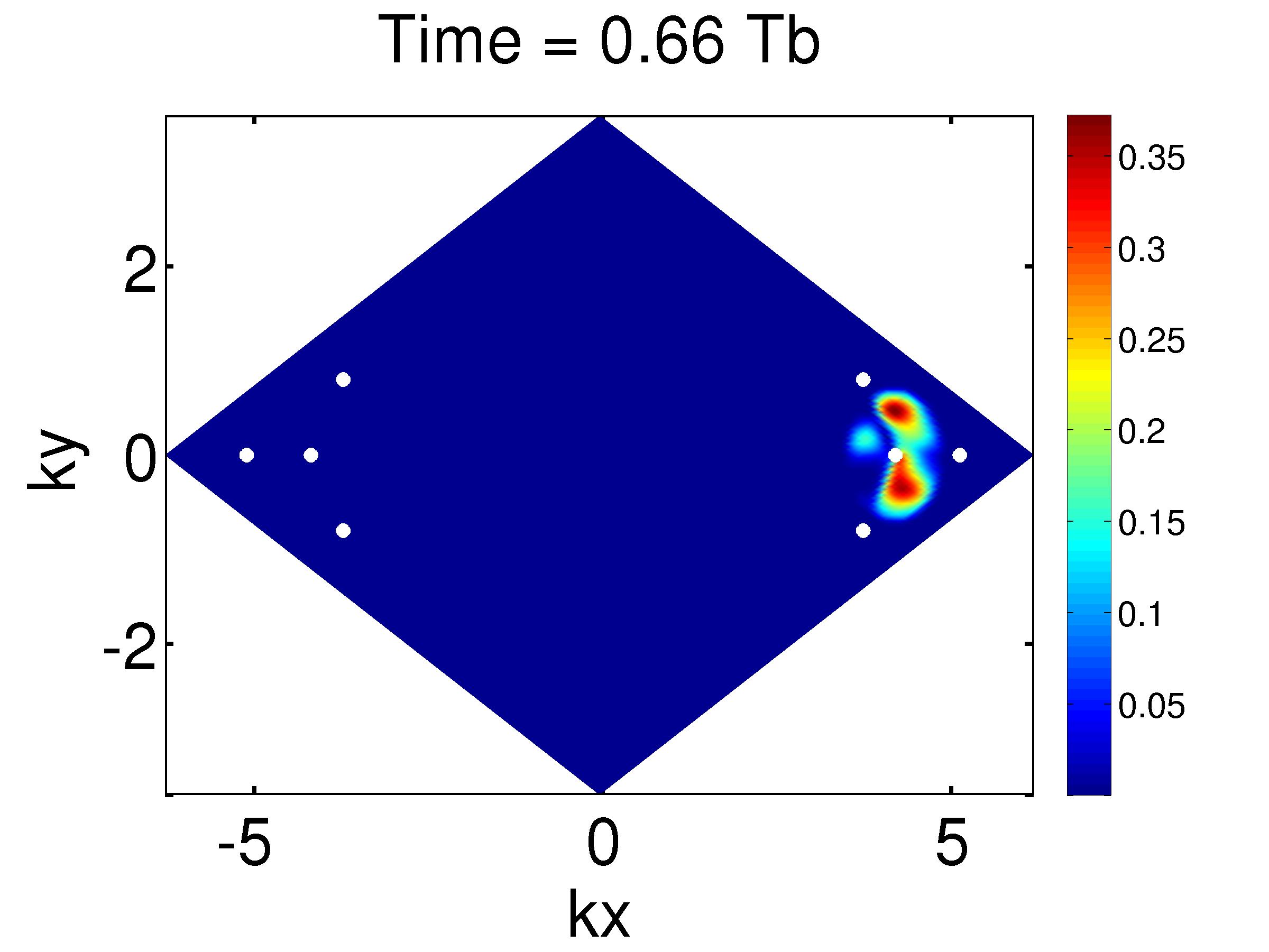}
 \caption{(color online) The quasi-momentum distribution of the second band for Bloch-Zener oscillations as a function of $(k_x,k_y)$ at $t' = 0.5t$ resulting from a force acting along the $\hat{e}_1 + \hat{e}_2 = \hat{x}$ direction. Total number of particles considered is 256.}
 \label{fig:kh_050_xpyp}
 \end{center}
 \end{figure*}

\begin{figure}[ht]
 \begin{center}
\includegraphics[width= 4.2cm]{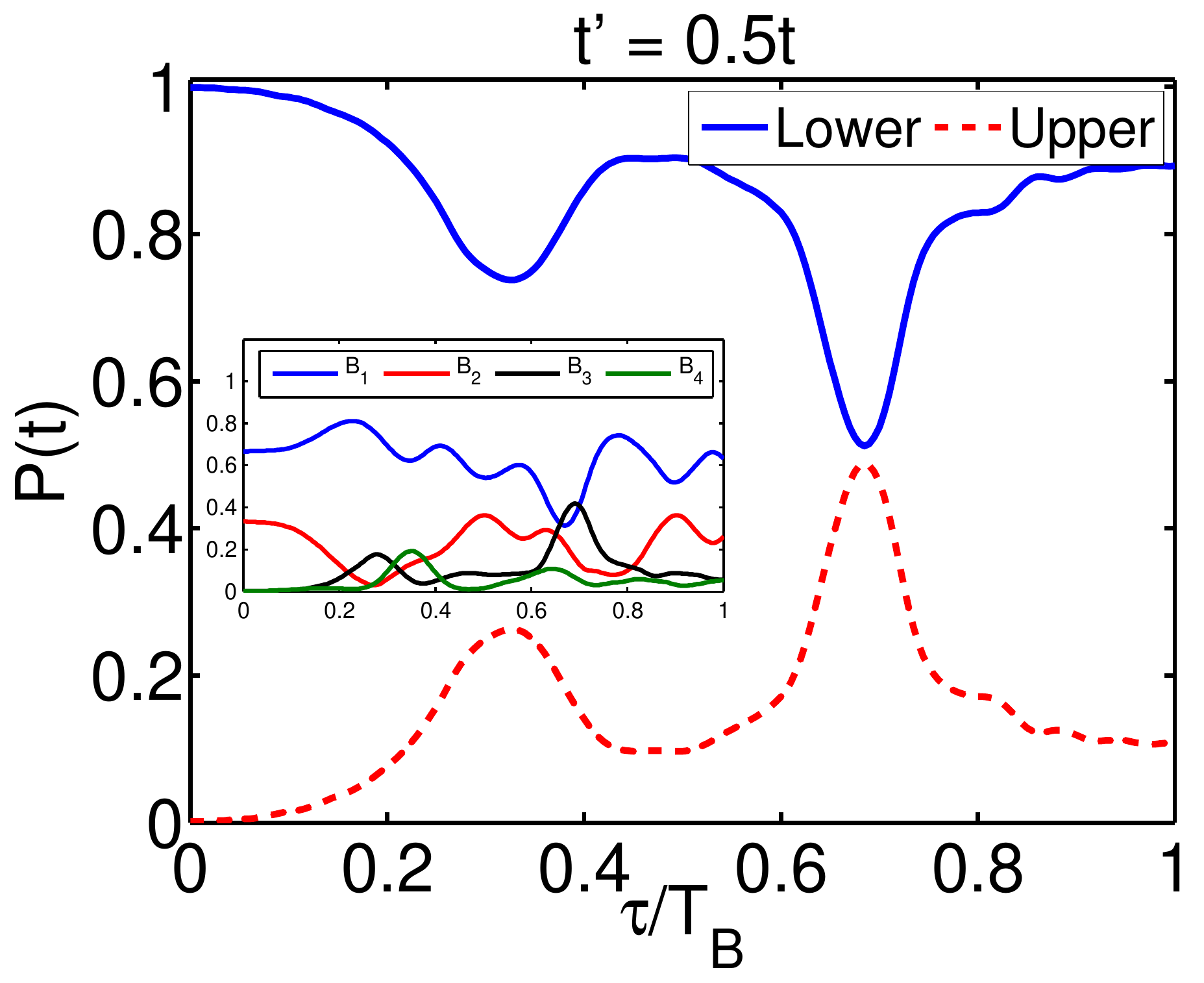}
 \includegraphics[width= 4.2cm]{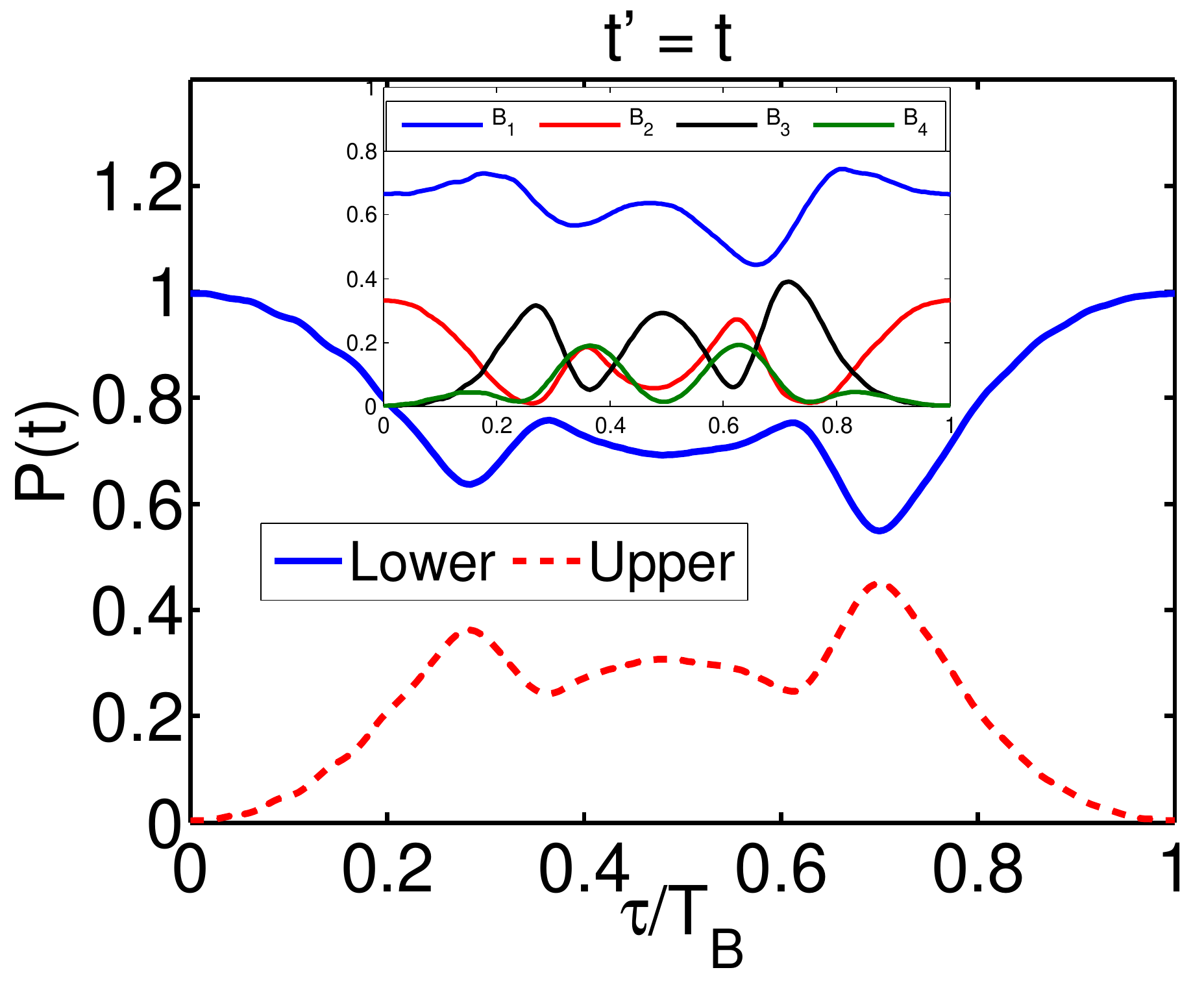}
 \caption{(color online) $P_l(\tau)$ is plotted in blue and $P_u(\tau)$ is plotted in red for the parameters in FIG.\eqref{fig:kh_050_xpyp}. In the inset we have the individual amplitudes for all the bands.}
 \label{fig:occprob_xpyp}
 \end{center}
 \end{figure}

In contrast with the above situation, we show in FIG.\eqref{fig:kh_050_xpyp} the Bloch-Zener oscillation of $256$ particles at $t' = 0.5t$ keeping the rest of the parameters unchanged. At this $t'$ and in this direction the system has four DPs, all of which are probed at different times by the cloud. There is a transfer of amplitude when the state passes through each of the four DPs. 

In FIG.\eqref{fig:occprob_xpyp} we plot $P_l(\tau)$ and $P_u(\tau)$ for $t' = 0.5t$. The lower band only shows two peaks even though this system has four DPs. This is because we have been unable to resolve the transfer at each of the four DPs to high accuracy which is obtained only when there is a reasonable fraction of the cloud in the second band at time $\tau = 0$. This requires a large number of particles in the cloud, increasing its width in momentum space. This decreases the resolution of the DPs, and can only be circumvented by increasing the size of the system, which is limited by currently available computational power to us.

No DPs are encountered when we apply the force field in the $\hat{e}_1 = \frac{1}{2}\hat{x} - \frac{\sqrt{3}}{2}\hat{y}$ direction or the $\hat{e}_2 = \frac{1}{2}\hat{x} +\frac{\sqrt{3}}{2}\hat{y} $ direction. The quasi-momentum distribution for the second band corresponds to the second Brillouin zone in the momentum distribution obtained in optical lattice experiments. Thus from the above discussion, we see that the DPs can be probed in such experiments.

\subsubsection{Signal for PB curvature: Rotating condensates}\label{sec:theoryrotate}\label{sec:curv_kh_model}
In this section we propose a method of measuring the PB curvature in optical lattice experiments different from methods proposed previously \cite{price2012, atala2012}. The hamiltonian in optical lattice experiment is generated as given in Eq. \eqref{opticalhami}. Then using time of flight experiments which are the standard methods of imaging in optical lattice experiments, we study the density profiles of the atomic cloud. Here all the external potentials are switched off suddenly following which the atomic cloud is allowed to expand ballistically, the atoms now behaving as free particles. Images of the cloud at various intervals of time shows how the density distribution changes as the cloud expands. If the system initially had non-zero PB curvature, then the cloud along with expansion, rotates thereby probing the PB curvature.
 
To understand the density profiles we first review the Sundaram-Niu\cite{sundaram1999} equations (SNE) and then the Thomas-Fermi approximation for a many-fermion system to address the problem of rotating condensates. 

The Sundaram-Niu equations \cite{sundaram1999} govern the classical dynamics of a wave packet restricted to the band with momentum space width small compared to that of the Brillioun zone and a real space width small compared to the applied external field. These wave packets therefore have a width in real space that is large compared to the lattice spacing but small compared to the scale of the variation of the external fields. The equation of motion of the Bloch electron for two dimensional systems in the absence of magnetic field are
\begin{eqnarray}
\label{gneq1}
\dot x^i&=&\frac{1}{\hbar}\frac{\partial\epsilon(k)}{\partial k_i}
+{\cal B}(k)\epsilon^{ij}\dot k_j\\
\label{gneq2}
\hbar \dot k_i&=&-\frac{\partial V(x)}{\partial x^i}
\end{eqnarray}
where $\epsilon(k)$ is the energy in the absence of external fields, ${\cal B}(k)$ is the PB field and $V$ the external potential. Thus ${\cal B}(k)$ induces a force-dependent anomalous velocity. 

The Sundaram-Niu equations describe the wave-packet dynamics for a single particle. In optical lattice experiments, the many body dynamics of the atoms needs to be considered for which we use the Thomas-Fermi approximation. This approximation assumes that the ground state is described by a phase-space particle density that incorporates the Pauli exclusion principle. The number of fermions in a phase-space volume $d^2xd^2p$ around the point $(x,p)$, $\tilde\rho(x,p)$ can be written as
\begin{equation}
\label{psrho}
\tilde\rho(x,p)=\frac{1}{(2\pi\hbar)^2}\Theta(\epsilon_F-h(x,p)) 
\end{equation}
where $\epsilon_F$ is the Fermi energy level and $h(x,p)$ is the single particle hamiltonian given by
\begin{align}
h(x,p) & = \epsilon(p) + V(x).
\end{align}
The initial particle density of the cloud can be calculated from Eq.\eqref{psrho}. We now allow the cloud to expand freely and by using time of flight experiments the density at various times can be computed using Liouville's theorem.

 The Thomas-Fermi approximation can be extended to multiple bands if the bands are well separated and if the applied external potential varies slowly enough to prevent interband transitions. Thus, the total phase space density is the sum of the phase space densities for first band and second bands. Additionally the Thomas-Fermi approximation fails when there are DPs in the system. 

The above formalism can be applied to our multi-band model since the bands are well separated. A slowly varying external potential is applied and the inversion symmetry is broken to open up a gap at the DPs. We use the same parameters as in \cite{uehlinger2013}. The system is confined by a rotationally invariant external harmonic potential $V(R)$, where
\begin{align}
V(R) & = \gamma_xX^2+\gamma_yY^2
\end{align}
and $\gamma_x =\frac12 \{ m\omega_{xt}^2(\lambda/\sqrt{3})^{2} \}$, $\gamma_y =\frac12 \{ m\omega_{yt}^2(\lambda/\sqrt{3})^{2} \}$ and $X,Y$ are the dimensionless spatial coordinates of the lattice sites. Here $\lambda=1064$ nm is the wavelength of the laser beam, $\omega_{xt} = \omega_{yt} = 40\pi $rad/s is the trapping frequency of the potential, $m$ is the mass of the ${}^{40}K$ atoms and $t/h=580$ Hz is the nearest neighbour hopping parameter. A staggered mass term $W /h = 0.1$ Hz is added. 

%
 \begin{figure}[hb]
  \includegraphics[width=4.2cm]{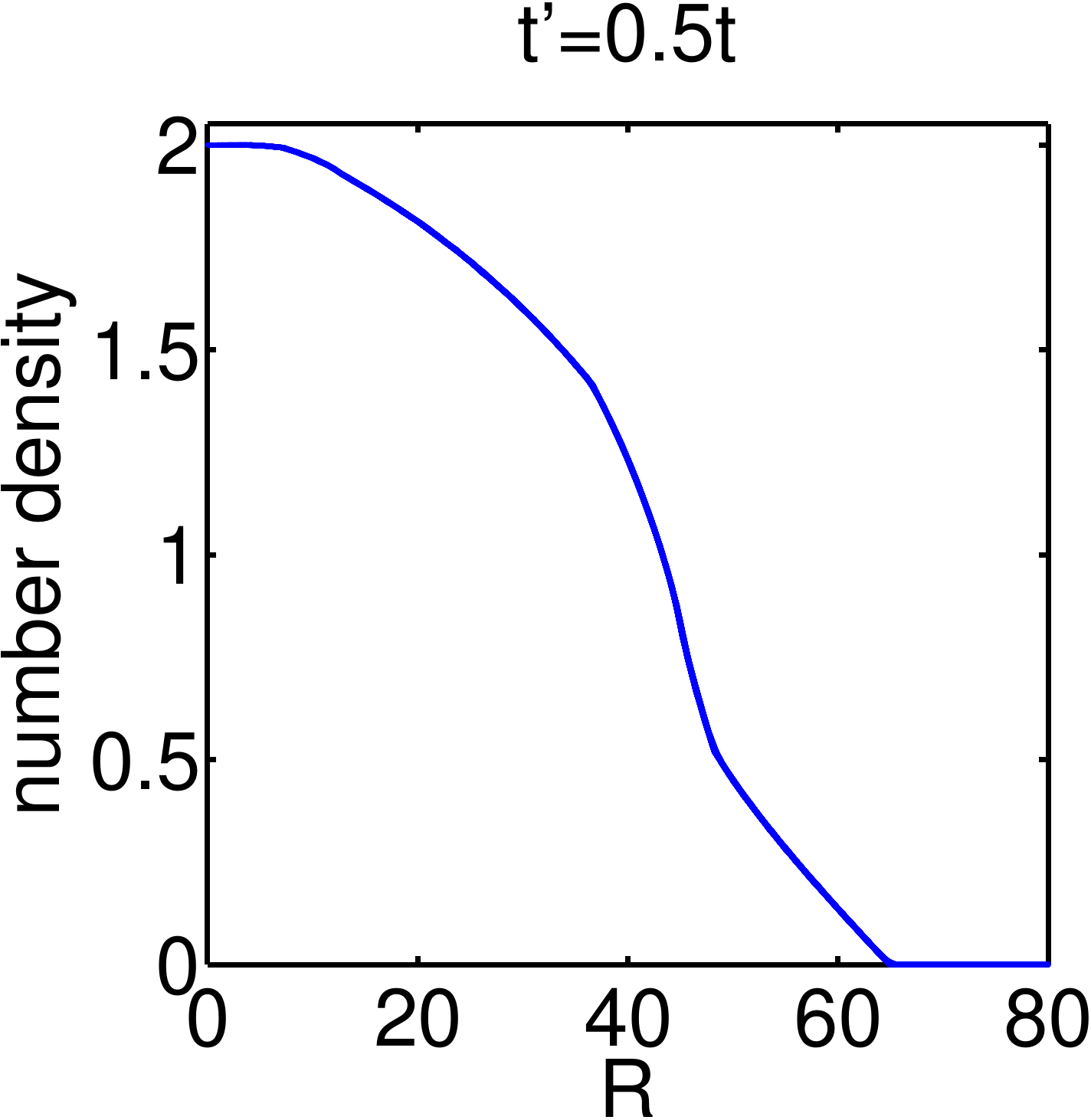}
  \includegraphics[width=4.2cm]{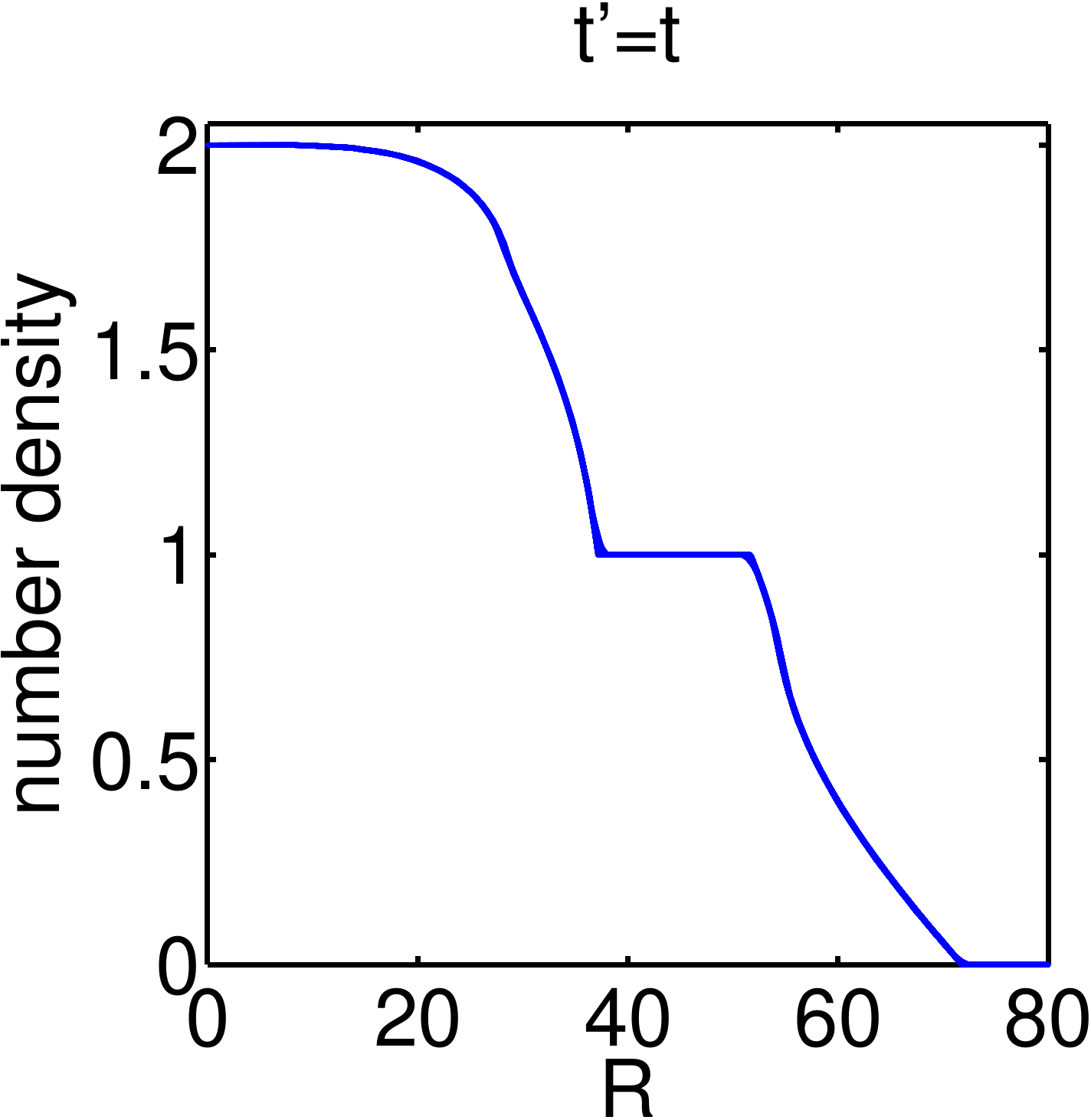}
  \caption{(color online) Particle density plot as a function of $R$ for $t' = 0.5t$ and $t' = t$.}
  \label{fig:density}
  \end{figure}

FIG.\eqref{fig:density} shows the variation of the number density of the cloud with distance $R$ for both $t'=0.5t$ and $t'=t$. At $R=0$ it reaches a maximum since the trap potential is zero there. Away from $R=0$, the total energy of each of the bands increases by $V(R)$, reducing the number of occupied states lying below $\mu = 0$ and thus decreasing the density. 
At $t' = t$ there is a plateau where the density has a constant value of $1$. The plateau occurs due to the energy gap between the second and the first band. The size of the plateau decreases depending on the size of the gap between the two bands. Below $t' = 0.717$, the first and the second bands overlap and hence we see a continuous change in the density as seen from the plot for $t' = 0.5t$.

\begin{figure}[ht]
  \includegraphics[width=4.2cm]{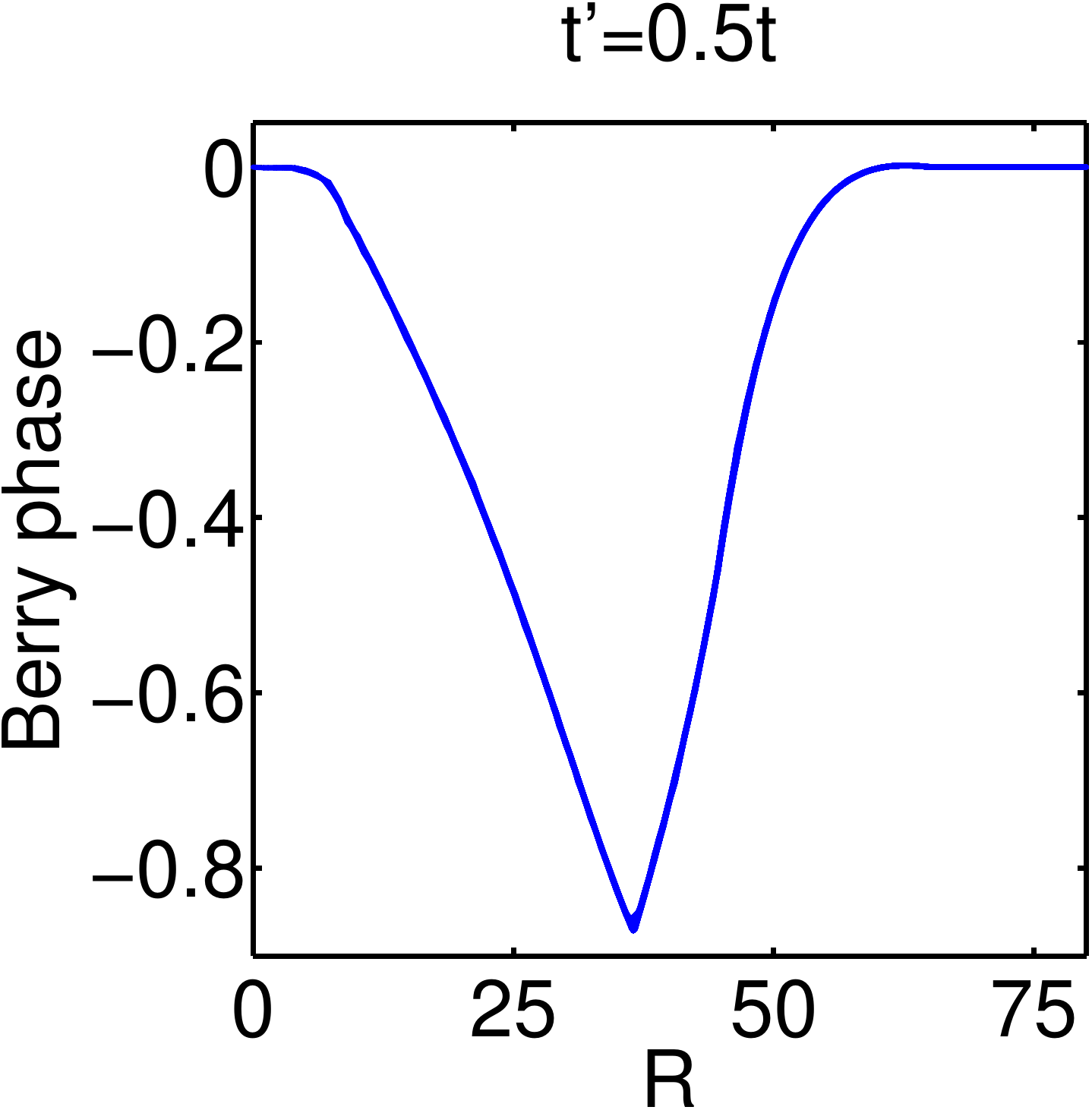}
  \includegraphics[width=4.2cm]{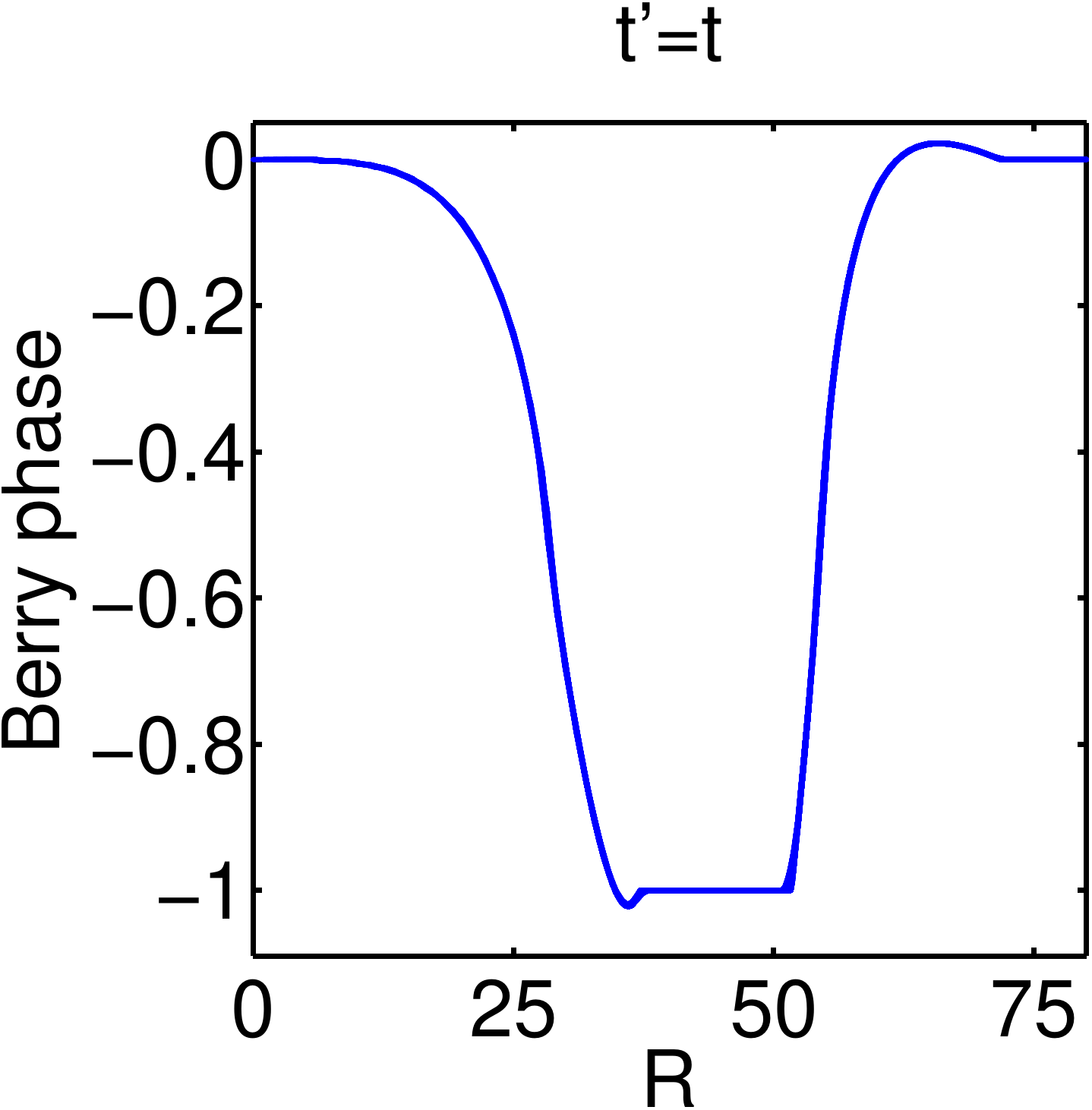}
  \caption{(color online) PB phase as a function of $R$ for $t'=0.5t$ and $t'=t$.}
  \label{fig:berryphase}
  \end{figure}

In FIG.\eqref{fig:berryphase}, we see that the PB phase of the occupied bands is zero at the center $R = 0$, since the Chern number of the contributing bands are equal and opposite. Away from the centre the trap potential reduces the number of occupied states below $\mu = 0$, decreasing the PB phase as a consequence. A minima is reached when the only contributing band is the lowest band after which the PB phase increases and reaches a value of zero when all the bands are empty. At $t' = t$, the PB phase also shows a plateau similar to that of the density whereas at $t' = 0.5t$ the PB phase changes continuously. At $t' = t$, there are kinks where the PB phase is lesser than $-1$ and greater than $0$. 
%
%
The kink appears because the PB phase of the band is not entirely positive when the Chern number is $+1$ and vice-versa. To illustrate this we plot, in FIG.\eqref{fig:berry_phase_filling}, the PB phase of first band as a function of the filling factor. 

\begin{figure}[!h]
  \includegraphics[width=4.2cm]{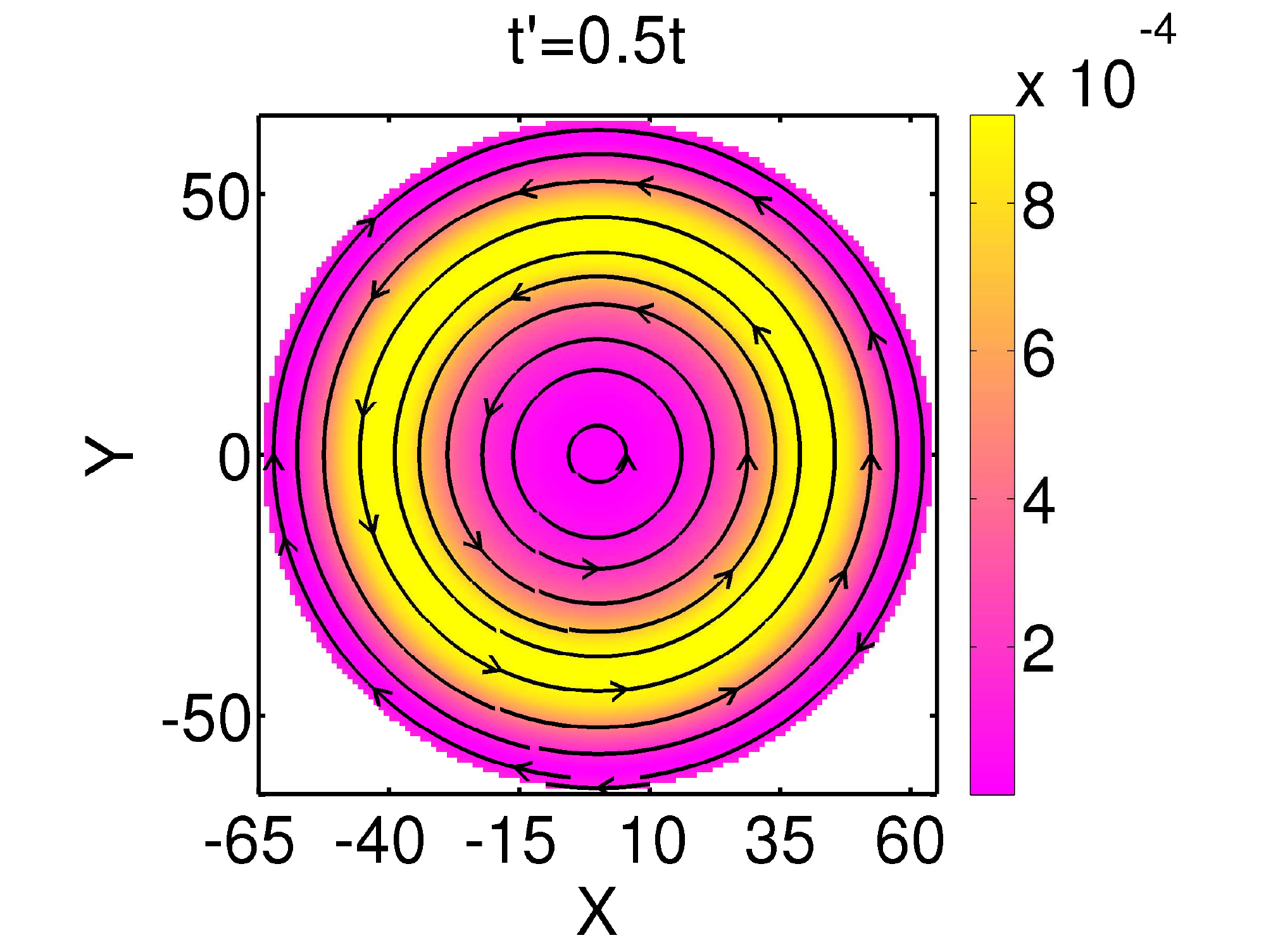}
  \includegraphics[width=4.2cm]{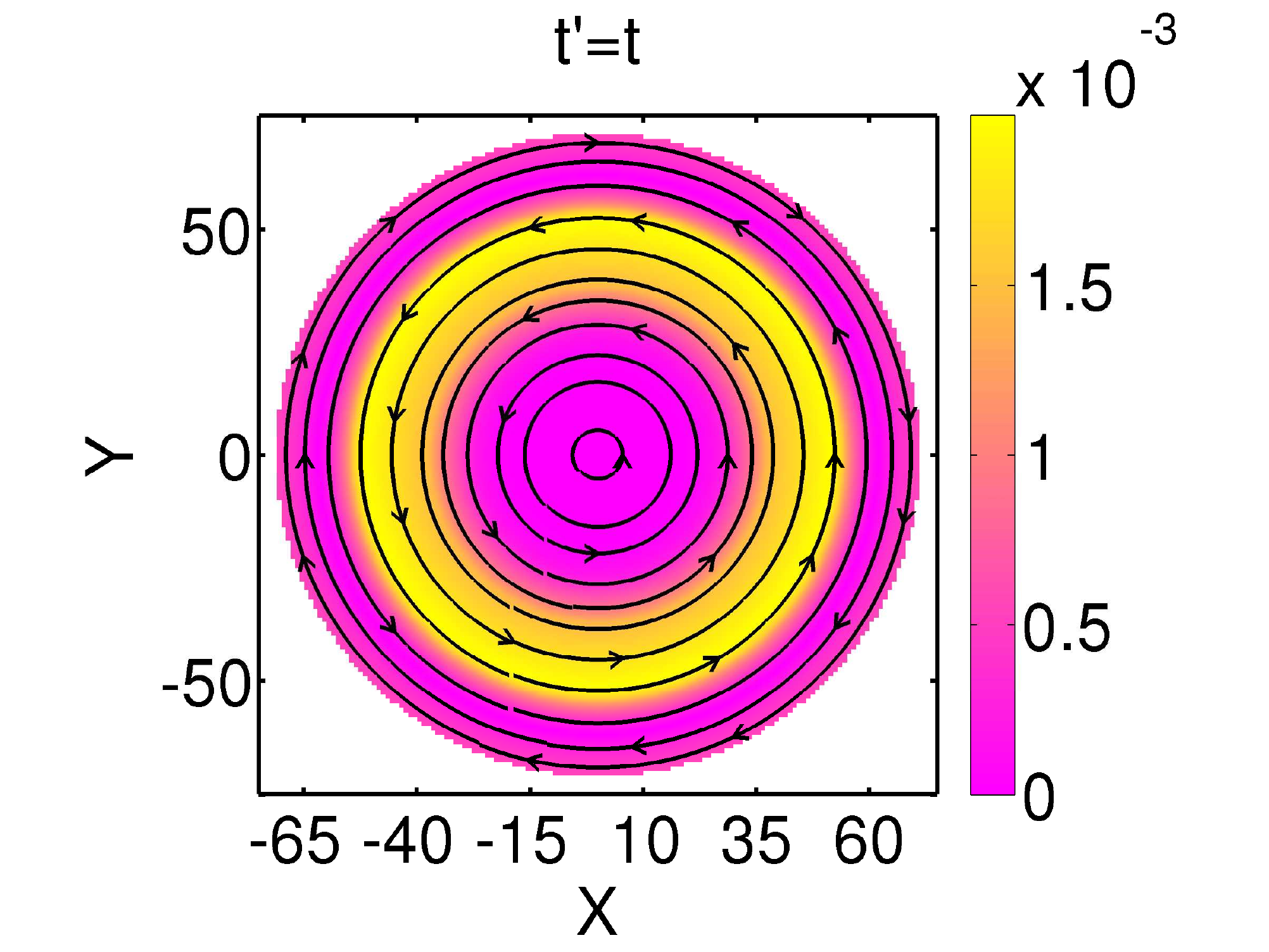}
  \caption{(color online) Velocity vector measured in m/s as a function of $X$ and $Y$ for $t'=0.5t$ and $t'=t$.}
  \label{fig:velovec}
  \end{figure}

From the Thomas Fermi approximation, the velocity of the cloud can be computed and is as shown in FIG.\eqref{fig:velovec}. The velocity is zero where the PB phase is zero. The direction of the velocity at the boundary changes due to the change in the sign of the PB phase as shown in FIG.\eqref{fig:berry_phase_filling}. 
From the velocity, the total angular momentum density per particle $L$ can be computed and is found to be $\approx 12\hbar$ for $t' = t$ and $\approx 6\hbar$ for $t' = 0.5 t$. This value is large compared to the value obtained in bosonic optical lattice experiments \cite{chevy2000}, and thus should be observable in such experiments.

 \begin{figure}[!h]
  \includegraphics[width=4.2cm]{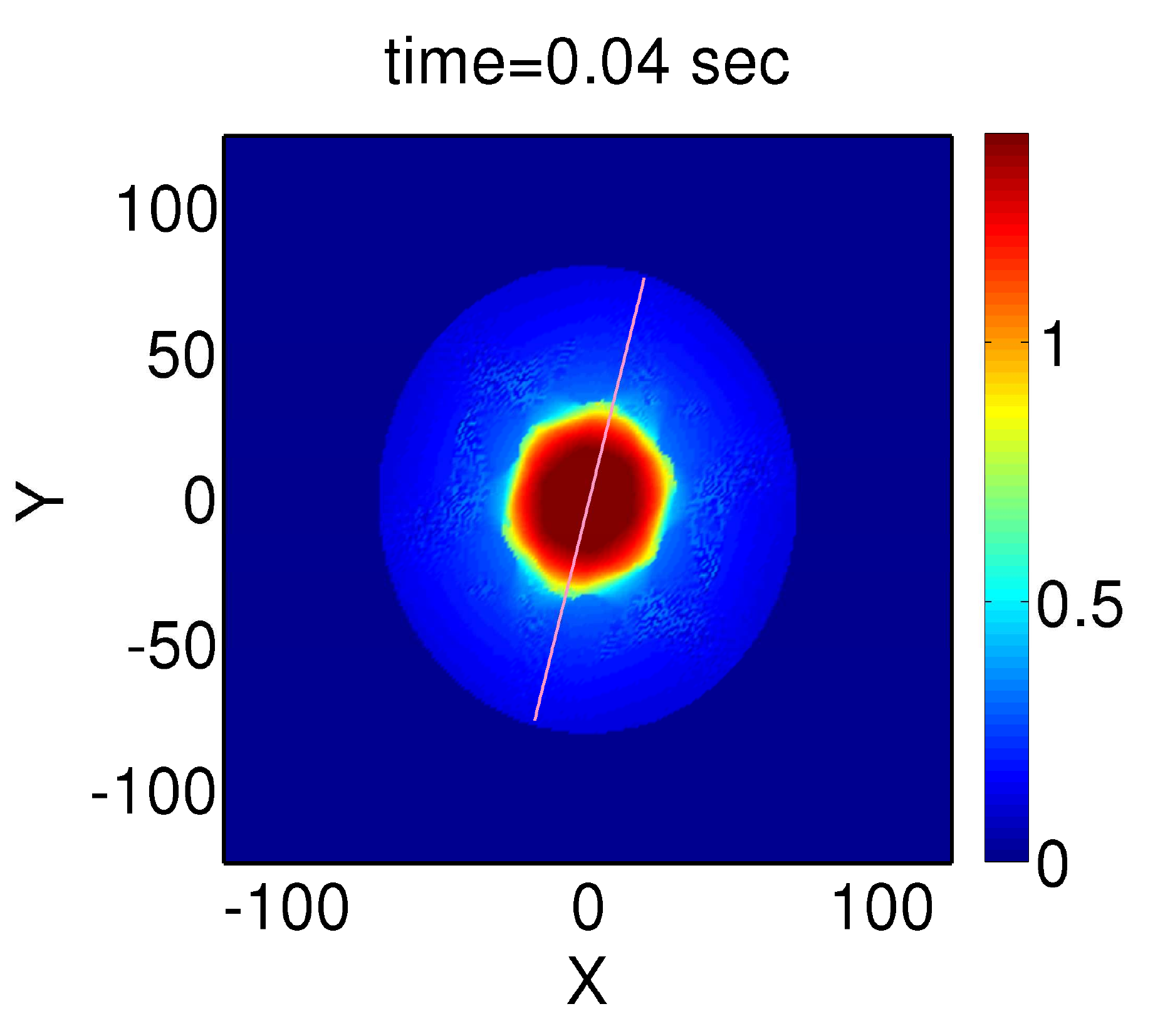}
  \includegraphics[width=4.2cm]{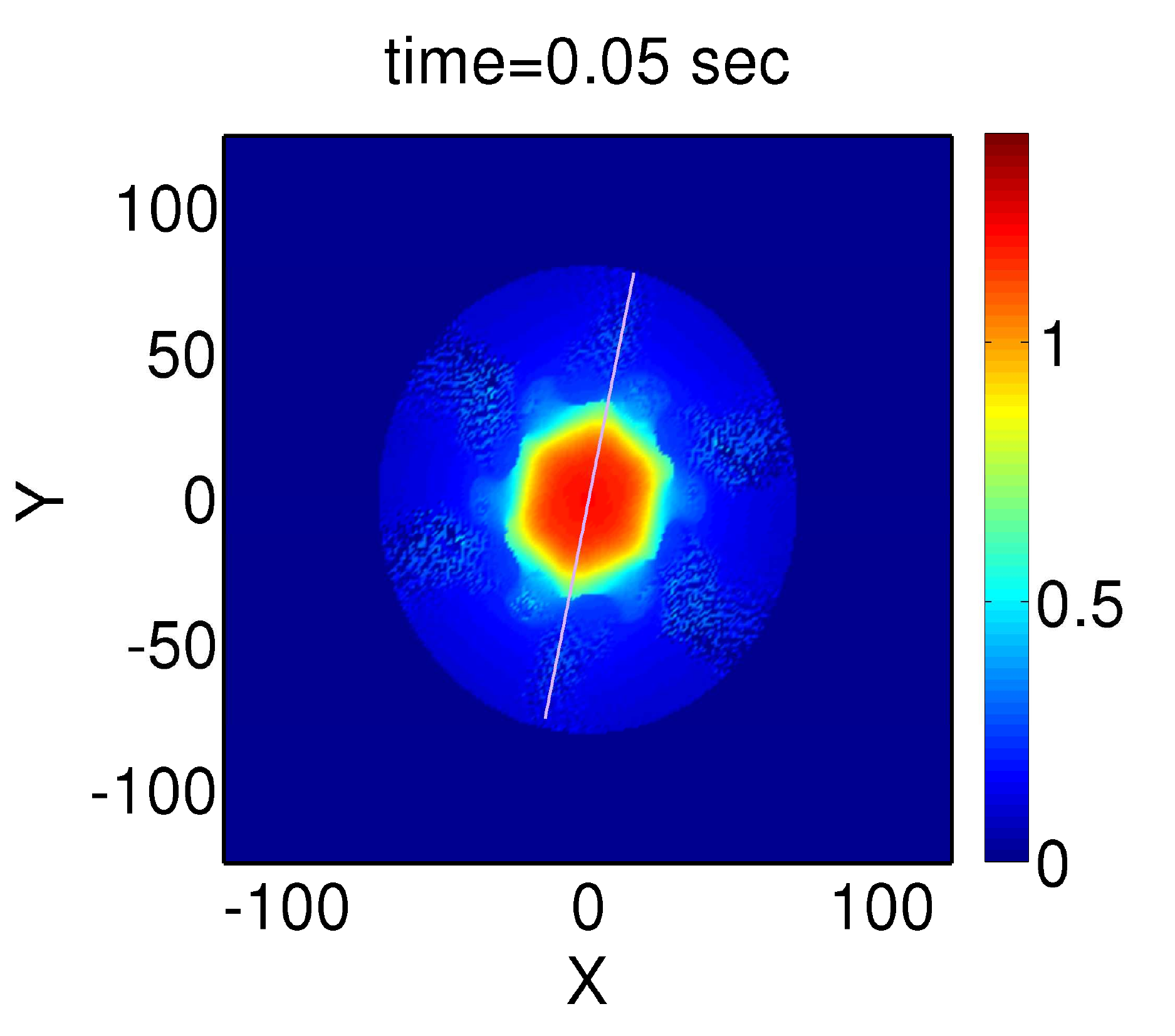}
  \includegraphics[width=4.2cm]{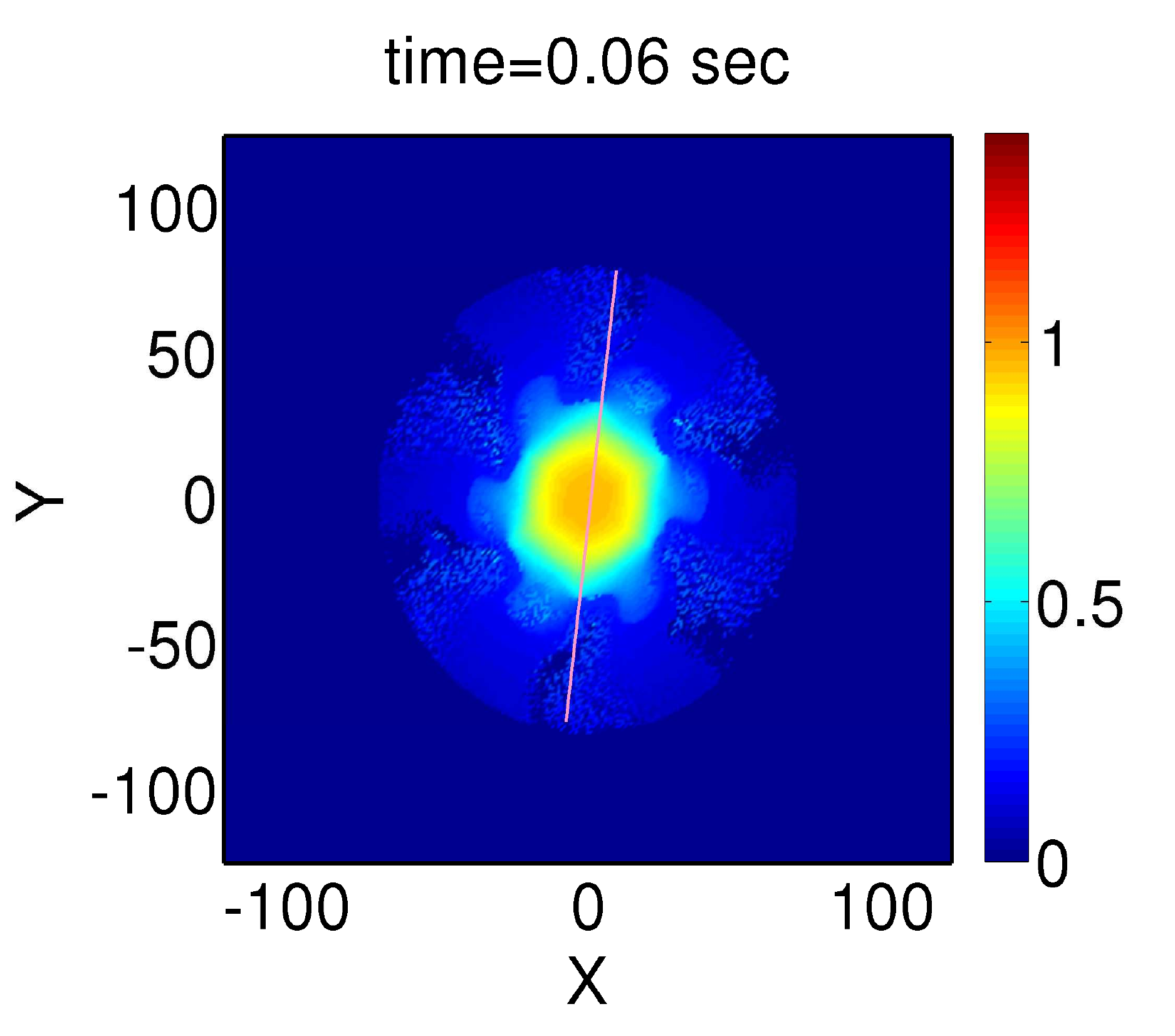}
  \includegraphics[width=4.2cm]{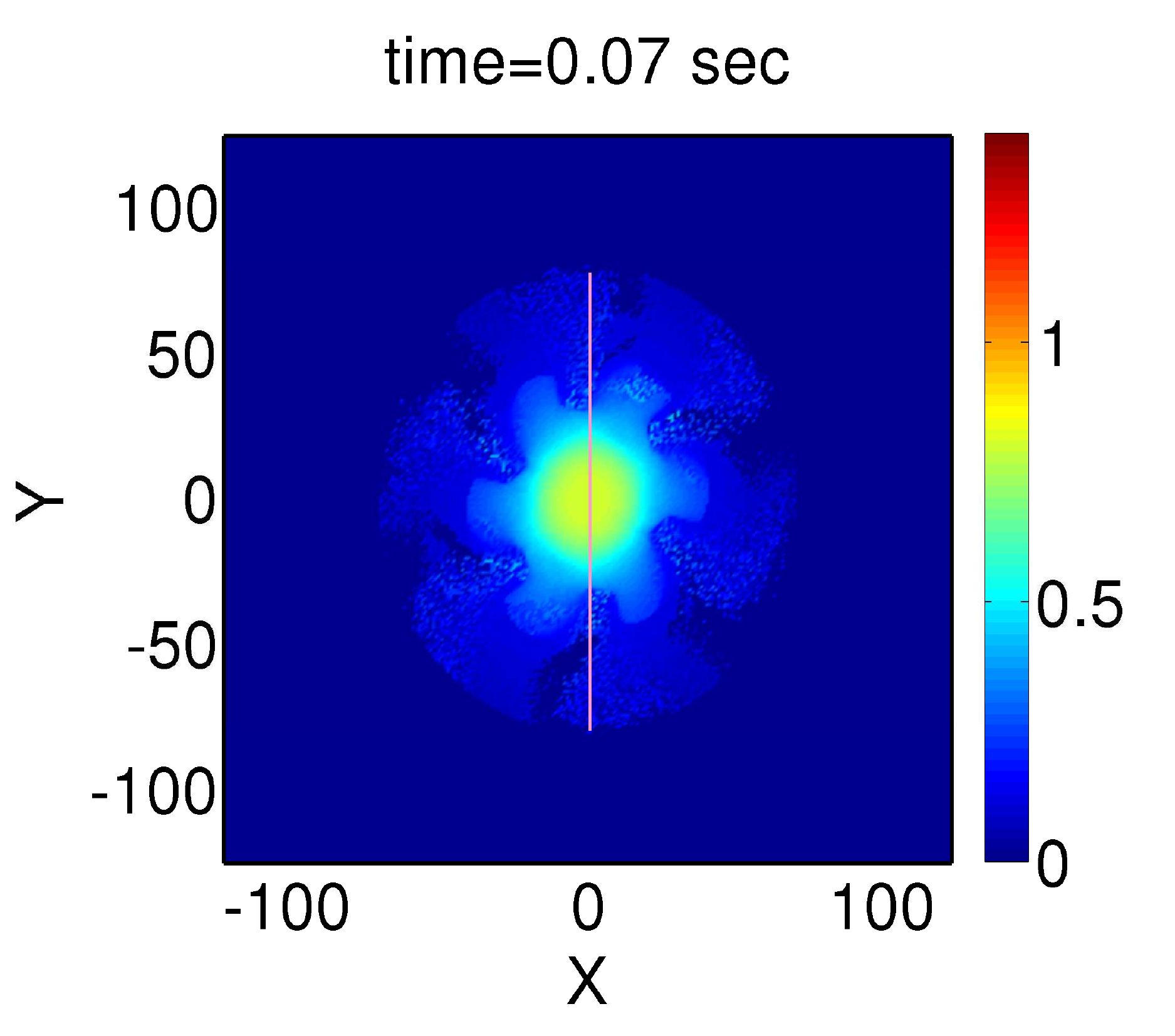}
  \caption{(color online) Density profiles as a function of $X,Y$ for $t'=t$ at different times. Lines are drawn in the above figures to show the rotation of the cloud as a function of time due to non-zero PB curvatures.}
  \label{fig:rotation}
 \end{figure}


The evolution of the cloud density after the traps have been switched off can be computed using Liouville's theorem. At various times the cloud rotates as it expands as seen in FIG.\eqref{fig:rotation}. This rotation strongly indicates a non-zero PB curvature. The cloud also inherits the hexagonal structure of the underlying  honeycomb lattice as it expands.

\section{Conclusions and Discussion}
To summarize, we have analyzed a fermionic model on the honeycomb lattice with spin-dependent hopping that breaks time-reversal symmetry but preserves the particle-hole (chiral/sub-lattice) symmetry, the Kitaev-Hubbard model. The model has DPs and is a semi-metal at half filling. We show that the particle-hole symmetry implies that the total PB curvature vanishes everywhere on the Brillioun zone at half filling.  Consequently, the generalized Zak phases are topological invariants that are wholly determined by the positions and chiralities of the DPs. We show that all this information about the topology is contained in the determinant of a matrix $\Sigma(k)$ defined in Eq.\eqref{sigmadef}.  We also numerically show that the structure of the non-dispersive edge states are determined by these topological invariants.

Multiple DPs exist in this model and as the strength of the spin-dependent hopping parameter, $t'$, is varied, topological Lifshitz transitions occur where the DPs are created and merge. At $t'=0$, the model is same as graphene and there are 4 DPs. As soon as $t'$ changes from zero, there is a transition from $4$ to $8$ DPs.  As $t'$ increases the DPs migrate over the Brillioun zone and pairwise merge at $(0\pi)$, $(\pi,0)$ and $(\pi,\pi)$, resulting in a transition $2$ DPs at $t'=1/\sqrt{3}$.  At $t'>\sqrt{3}$, there is again a transition to $8$ DPs which now emerge from $(0,0)$.  The signal of these transitions can be seen in the density of states and in the edge state structure. The effect of broken time reversal symmetry of this model can be seen from the existence of the charge currents at the edges of the system. We observe that the edge currents change sign near the transitions.

A scheme to realise spin-dependent hopping in cold atom systems was proposed by Duan {\em et al} \cite{duan2003}. We have analyzed their scheme and have provided a systematic derivation of the low energy effective spin-dependent periodic potential which leads to the spin-dependent hopping matrix elements in the tight-binding model.

Finally, we have examined experimental signals of the topological and geometric features of the model realized in cold atom systems. We have shown that Bloch-Zener oscillations in our system probes the location of the DPs and can hence be used to observe the creation, migration and the merging process. The confining trap breaks the particle-hole symmetry and makes the PB curvatures of the bands observable. We have shown that the effect of this is seen in the rotation of the expanding cloud when the trap is removed. For realistic atomic and trap parameters, we have shown that this provides a clear signal in time of flight experiments. 

In conclusion, while the Kitaev-Hubbard model was initially proposed \cite{duan2003} to realize Kitaev's honeycomb model at large $U$ and half filling, our work \cite{hassan2013,hassan2013b} shows that it contains rich physics at all parameter ranges. At half filling, intermediate and large $U$, there are phases with magnetic order and a transition to an algebraic spin liquid \cite{hassan2013}. At small to intermediate $U$ it shows a very interesting phenomena of creation, migration and merging of DPs \cite{jeanpaul2013}. 

We have also shown that at quarter and three quarter filling, there is a Chern insulating phase which indicates that there could be fractional anomalous quantum Hall states when the bands are partially filled. We feel that all these theoretical results strongly motivate attempts to physically realize the model.

{\bf Acknowledgement}: We thank Mukul S. Laad and David S\'en\'echal for useful discussions.  
\bibliography{biblio}
\end{document}